\documentclass[twocolumn,preprintnumbers,amsmath,amssymb,superscriptaddress]{revtex4}
\usepackage{graphicx}% Include figure files
\usepackage{dcolumn}% Align table columns on decimal point
\usepackage{bm}% bold math
\usepackage{textcomp}
\usepackage{color}
\usepackage{natbib}
\usepackage{amsmath, amsfonts, amssymb}
\usepackage{enumerate}
\usepackage[breaklinks]{hyperref}
\usepackage[hyphenbreaks]{breakurl}
\usepackage{paralist}
\usepackage{etex}

\newcommand{\ket}[1]{\left| #1 \right\rangle}

\begin{document}

%\preprint{APS/123-QED}
\title{Interacting topological edge channels}
%\title{Interaction induced conductance quantization in a HgTe based topological quantum point contact}% Force line breaks with \\

\author{Jonas Strunz}
\altaffiliation{All three authors contributed equally to this work, email: Jonas.Strunz@physik.uni-wuerzburg.de}
\affiliation {Experimentelle Physik III, Physikalisches Institut, Universit\"at W\"urzburg, Am Hubland, D-97074 W\"urzburg, Germany}
\affiliation{Institute for Topological Insulators, Am Hubland, D-97074 W\"urzburg, Germany}

\author{Jonas Wiedenmann}
\altaffiliation{All three authors contributed equally to this work, email: Jonas.Strunz@physik.uni-wuerzburg.de}
\affiliation {Experimentelle Physik III, Physikalisches Institut, Universit\"at W\"urzburg, Am Hubland, D-97074 W\"urzburg, Germany}
\affiliation{Institute for Topological Insulators, Am Hubland, D-97074 W\"urzburg, Germany}

\author{Christoph Fleckenstein}%
\altaffiliation{All three authors contributed equally to this work, email: Jonas.Strunz@physik.uni-wuerzburg.de}
\affiliation{%
Institute of Theoretical Physics and Astrophysics, University of W\"urzburg, 97074 W\"urzburg, Germany}%

\author{Lukas Lunczer}
\affiliation {Experimentelle Physik III, Physikalisches Institut, Universit\"at W\"urzburg, Am Hubland, D-97074 W\"urzburg, Germany}
\affiliation{Institute for Topological Insulators, Am Hubland, D-97074 W\"urzburg, Germany}

\author{Wouter Beugeling}
\affiliation {Experimentelle Physik III, Physikalisches Institut, Universit\"at W\"urzburg, Am Hubland, D-97074 W\"urzburg, Germany}
\affiliation{Institute for Topological Insulators, Am Hubland, D-97074 W\"urzburg, Germany}

\author{Valentin L. M\"uller}
\affiliation {Experimentelle Physik III, Physikalisches Institut, Universit\"at W\"urzburg, Am Hubland, D-97074 W\"urzburg, Germany}
\affiliation{Institute for Topological Insulators, Am Hubland, D-97074 W\"urzburg, Germany}

\author{Pragya Shekhar}
\affiliation {Experimentelle Physik III, Physikalisches Institut, Universit\"at W\"urzburg, Am Hubland, D-97074 W\"urzburg, Germany}
\affiliation{Institute for Topological Insulators, Am Hubland, D-97074 W\"urzburg, Germany}

\author{Niccol\'o Traverso Ziani}
\affiliation {Institute of Theoretical Physics and Astrophysics, University of W\"urzburg, 97074 W\"urzburg, Germany}
\affiliation {Dipartimento di Fisica, Universit\`a di Genova, CNR-SPIN, Via Dodecaneso 33, 16146 Genova, Italy}

\author{Saquib Shamim}
\affiliation {Experimentelle Physik III, Physikalisches Institut, Universit\"at W\"urzburg, Am Hubland, D-97074 W\"urzburg, Germany}
\affiliation{Institute for Topological Insulators, Am Hubland, D-97074 W\"urzburg, Germany}

\author{Johannes Kleinlein}
\affiliation {Experimentelle Physik III, Physikalisches Institut, Universit\"at W\"urzburg, Am Hubland, D-97074 W\"urzburg, Germany}
\affiliation{Institute for Topological Insulators, Am Hubland, D-97074 W\"urzburg, Germany}

\author{Hartmut Buhmann}
\affiliation {Experimentelle Physik III, Physikalisches Institut, Universit\"at W\"urzburg, Am Hubland, D-97074 W\"urzburg, Germany}
\affiliation{Institute for Topological Insulators, Am Hubland, D-97074 W\"urzburg, Germany}

\author{Bj\"orn Trauzettel}
\affiliation{Institute of Theoretical Physics and Astrophysics, University of W\"urzburg, 97074 W\"urzburg, Germany}
\affiliation{W\"urzburg-Dresden Cluster of Excellence \textit{ct.qmat}, Germany}

\author{Laurens W. Molenkamp}
 \affiliation {Experimentelle Physik III, Physikalisches Institut, Universit\"at W\"urzburg, Am Hubland, D-97074 W\"urzburg, Germany}
 \affiliation{Institute for Topological Insulators, Am Hubland, D-97074 W\"urzburg, Germany}
\affiliation{W\"urzburg-Dresden Cluster of Excellence \textit{ct.qmat}, Germany}

\date{\today}% It is always \today, today,
\maketitle

%\tableofcontents

\textbf{Electrical currents in a quantum spin Hall insulator are confined to the boundary of the system. The charge carriers can be described as massless relativistic particles, whose spin and momentum are coupled to each other. While the helical character of those states is by now well established experimentally, it is a fundamental open question how those edge states interact with each other when brought in spatial proximity. We employ a topological quantum point contact to guide edge channels from opposite sides into a quasi-one-dimensional constriction, based on inverted HgTe quantum wells. Apart from the expected quantization in integer steps of $2 e^2/h$, we find a surprising additional plateau at $e^2/h$. We explain our observation by combining band structure calculations and repulsive electron-electron interaction effects captured within the Tomonaga-Luttinger liquid model. The present results may have direct implications for the study of one-dimensional helical electron quantum optics, Majorana- and potentially para-fermions.}\\
%alternative approach could be 1D, history of 1D systems
The quantum spin Hall effect has been predicted in several systems \cite{Haldane1988,Kane2005f,Kane2005a,Bernevig2006} and was first realized in HgCdTe/HgTe quantum wells \cite{Konig2007b}. Later, this phase was observed in other material systems such as InAs/GaSb double quantum wells \cite{Knez2011} and in monolayers of WTe$_2$ and bismuthene \cite{Wu2018, Reis2017}.
%The quantum spin Hall insulator (QSHI) state was first realized in HgTe quantum wells (QW) of appropriate thickness \cite{Bernevig2006, Kane2005f}.
The defining properties of this state, related to its helical nature, are well established by numerous experiments such as the observation of conductance quantization of two spin polarized edge channels $G_0=2e^2/h$ with $e$ the electron charge and $h$ the Planck's constant \cite{Konig2007b}. Additionally, non-local edge transport and spin-polarization of the edge channels were demonstrated by suitable transport experiments \cite{Roth2009, Brune2012}. We instead target a still open question, namely how helical edge states interact with each other.

A quantum point contact (QPC) can be used to guide edge channels from opposite boundaries of the sample into a constriction. Such a device allows for studies of charge and spin transfer mechanisms by, \emph{e.g.}, adjusting the overlap of the edge states %and helps understanding backscattering mechanisms of helical liquids
\cite{Hou2009a,Stroem2009,Teo2009,Dolcini2010,Krueckl2011,Zhang2011,Orth2013,Sternativo2014,Dolcini2015d,Papaj2016}.
 Besides the general interest in the study of transport processes in such a device, the appropriate model to describe the essential physics and to capture interaction effects of helical edge states is still unclear. The one-dimensionality of the helical edge modes suggests a description in terms of the Tomonaga-Luttinger liquid when electron-electron interactions are taken into account. %\cite{Giamarchi2003}.
In this respect, the QPC setup provides an illuminating platform as it may give rise to particular backscattering processes.

%Bernevig, Hughes and Zhang derived the archetypal BHZ-model \cite{Bernevig2006} by considering the low energy approximation of the HgTe quantum well (QW) band structure close to the inversion point (around the critical QW thickness $d_c \approx 6.3\,$nm). This model was successfully used to predict the topological transition in HgTe/HgCdTe QWs and is widely used to study properties of helical edge channel.
%{\color{blue}  With increasing QW thickness, the system remains in the topological phase with counterpropagating edge modes, however, details of the bandstructre are modified. In particular, the band crossing point (Dirac point) of the edge channel is not oriented in band gap anymore \cite{Skolasinski2017}.}
%The predictive power of the BHZ model quickly breaks down for thicker quantum wells, where the crossing of the edge channels (Dirac point) is not in the band gap anymore \cite{Skolasinski2017}.
%An alternative is given by the (helical) Tomonaga-Luttinger model, which is able to properly describe the electron-electron correlations and inter-edge scattering mechanisms relevant in a real 1D system \cite{Haldane1981, Kane1992, xx} \textit{@Christoph please reference the relevant papers in the field.}.
We present the realization of a QPC based on HgTe quantum wells as evidenced by the observation of the expected conductance steps in integer values of $G_0$. The newly developed lithographic process allows the fabrication of sophisticated nanostructures based on topological materials without lowering the material quality. It thus opens the path to conduct experiments of topological materials on mesoscopic scales important for the coherent control of helical edge channels and topological quantum computing.
Depending on the QPC width $W_{\textrm{QPC}}$ and quantum well thickness $d_{\textrm{QW}}$, we observe a fractional plateau at $0.5G_0$ in absence of an applied magnetic field. We label this phenomenon the \textit{0.5 anomaly} in resemblance to the 0.7 anomaly frequently observed in point contacts fabricated in more conventional semiconductors \cite{Micolich2011}. Self-consistent $k\cdot p$ calculations allow us to identify the most plausible transport mechanism. %The helical Tomonaga-Luttinger model is used to explain the fractional plateau
Using the theory of helical Tomonaga-Luttinger liquids, we associate the experimental results with the presence of a spin-gap. Bias and temperature dependencies of the $0.5$ anomaly are in agreement with such a gap. Furthermore, we identify an indicator of the conventional 0.7 anomaly in our devices when increasing the applied bias voltage. This observation is in qualitative agreement with the present theory and the explanation given for the 0.7 anomaly in Ref.~\cite{Bauer2013}.

\section{Realization of a Quantum spin Hall quantum point contact}
\begin{figure*}
\includegraphics[width=0.7\textwidth]{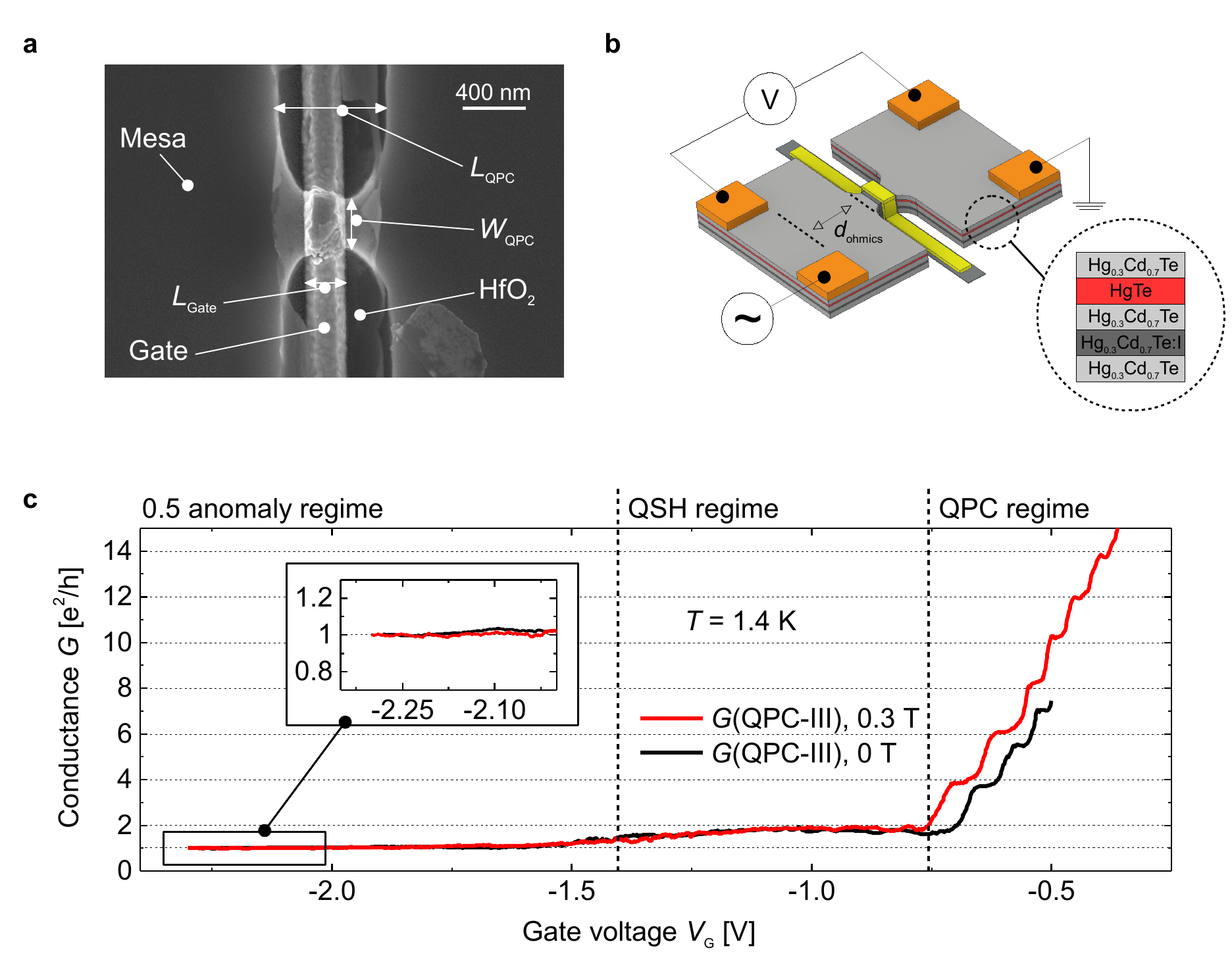}
\caption{\textbf{Realization of a topological quantum point contact:} \textbf{a}, Scanning electron micrograph of an exemplary device. A narrow channel is defined in the HgTe mesa with width $W_{\textrm{QPC}}$ and length $L_{\textrm{QPC}}$. A metallic gate electrode is separated from the mesa by a HfO$_2$ dielectric. \textbf{b}, Schematic of the QPC design and measurement setup. The gate electrode is depicted in yellow and the ohmic contacts in orange. The inset shows the epitaxially grown layer stack on a commercially available Cd$_{0.96}$Zn$_{0.04}$Te substrate. \textbf{c}, Gate voltage dependence of the conductance of QPC-III measured at $1.4\,$K. The conductance is divided into three regimes indicated by the vertical dashed lines. In the QPC regime, integer steps up to $14 e^2/h\,$ are observed. The pure quantum spin Hall regime is defined by a conductance of $2e^2/h$ (abbreviated as QSH regime in \textbf{c}). In the 0.5 anomaly regime an interaction driven gap opens leading to a quantized conductance of $e^2/h$. The inset shows a zoom of the $0.5$ anomaly regime.
}
\label{Fig1}
\end{figure*}

Figure~\ref{Fig1}a shows a scanning electron micrograph picture of a HgTe QPC. A constriction is formed by wet chemical etching of the HgTe heterostructure \cite{Bendias2018} and a top gate electrode is used to tune the chemical potential \cite{Kristensen1998}. The commonly employed approach of defining the QPC purely by electrostatic gating \cite{Wees1988} is not suitable in our case due to the presence of gapless edge modes with linear dispersion (Klein tunnelling) \cite{Katsnelson2006}. % Similar problems have been reported in materials with comparable band structure like graphene \cite{Terres2016, Tombros2011}.

Our devices are fabricated from HgTe quantum wells epitaxially grown on Cd$_{0.96}$Zn$_{0.04}$Te substrates and sandwiched between Hg$_{0.3}$Cd$_{0.7}$Te barriers (see inset Fig.~\ref{Fig1}b). The thickness of the HgTe layer, if not explicitly stated otherwise, is $d_\mathrm{QW} = 10.5\,\mathrm{nm} $.
The width of the channel $W_{\textrm{QPC}}$ ranges between $25\,$ to $250\,$nm, while the length $L_{\textrm{QPC}}$ is kept constant around $500\,$nm. The length of the gate electrode $L_{\textrm{Gate}}$ is approximately $200\mbox{-}300\,$nm. As depicted in Fig.~\ref{Fig1}b, ohmic contacts are placed far away ($d_{\textrm{ohmics}} \approx 80\, $\textmu m) from the constriction to allow full energy relaxation in the HgTe leads and to avoid geometrical resonances. Details about the fabrication process, material parameters and measurement setup are presented in the supplementary information, Sec.~I.

The conductance $G$ of a representative QPC as a function of applied gate voltage $V_G$ is depicted in Fig.~\ref{Fig1}c. Three regimes can be identified. For gate voltages $V_G\geq -0.75\,$V, we observe conventional QPC behaviour. Conductance plateaus at integer multiples of $G_0$ are developed and the quality of quantization can be improved by applying a small magnetic field (shown in red). % In consequence quantized steps up to values of $14 e^2/h$ are visible -- highlighting the quality of the QPC. % From the temperature dependence of the plateaus (see supplementary ) we are able to estimate an energy splitting of the sub-bands $\Delta E\ \lesssim 4 k_B T=4.8\,$meV.
For gate voltages between $-0.75\,$V$\,>V_G>-1.2\,$V the point contact is in the quantum spin Hall regime. A long plateau around $G_{\textrm{0}}$ is assigned to two helical edge channels.
For still more negative gate voltages $V_G\leq-1.2\,$V, a step-like transition from $G_{\textrm{0}}$ to a long plateau at $0.5G_0$ is observed. The inset shows the remarkable precision of the quantization even at zero magnetic field. This observation constitutes the main finding of this work.

\section{The $0.5$ anomaly}
\begin{figure*}[hbtp]
\includegraphics[width=0.7\textwidth]{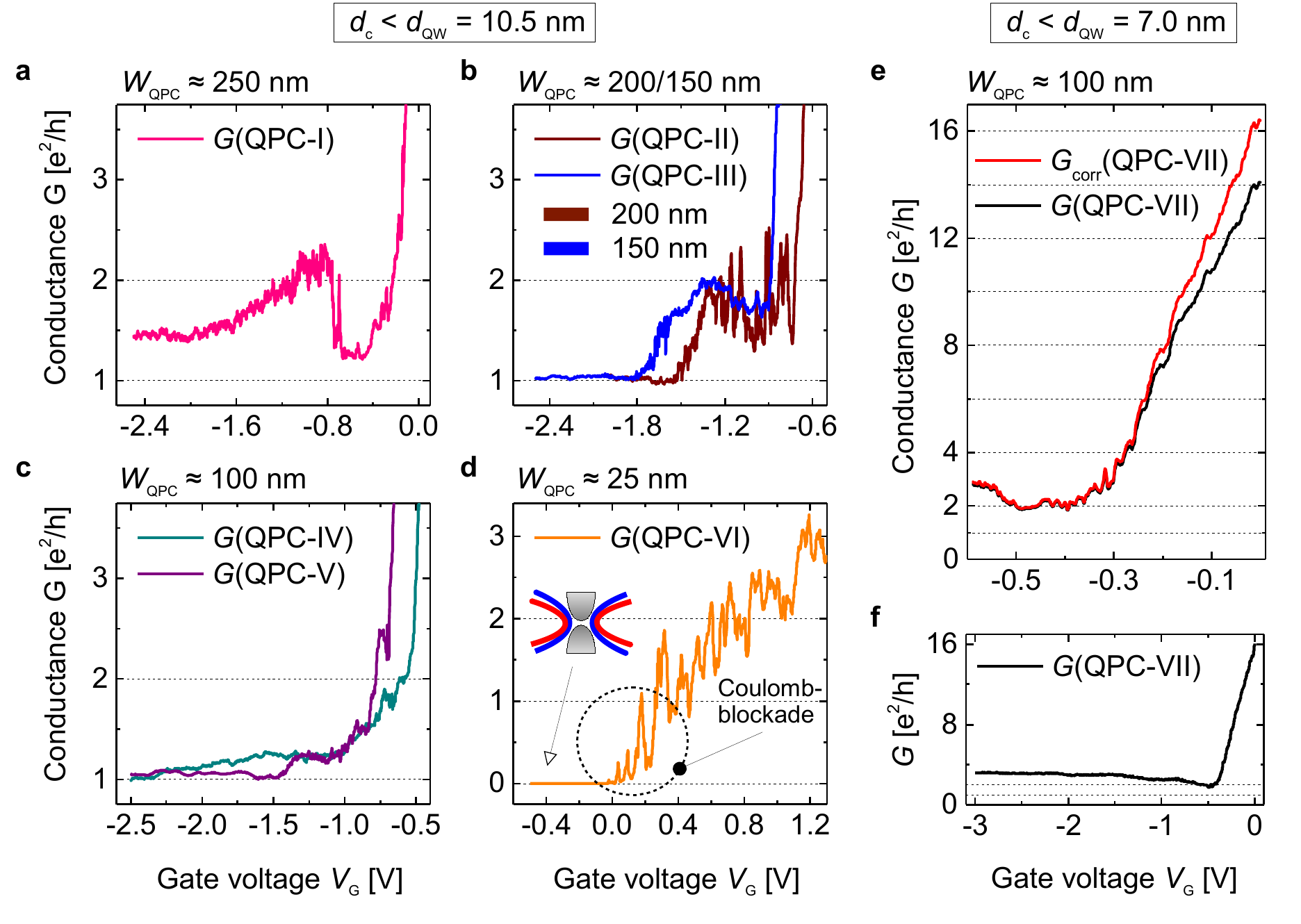}
\caption{\textbf{Width dependencies of the 0.5 anomaly:} \textbf{a-d}, Conductance as a function of gate voltage $V_{\textrm{G}}$ measured at zero magnetic field and a temperature of $T\approx 1.4\,$K for QPCs with varying width $W_{\textrm{QPC}}$ as indicated.
\textbf{e-f}, Conductance of a QPC based on a quantum well width $d_{\textrm{QW}}=7.0\,$nm. The raw data is depicted in black, a serial resistance of $260 \, \Omega$ was subtracted for the red graph. A wider gate voltage range is shown in the lower panel. }
\label{Fig2}
\end{figure*}

The $0.5$ anomaly is a robust signature. It is stable over multiple thermal cycles and we have reproduced it in several devices. An overview of various devices is presented in Fig.~\ref{Fig2}. The 0.5 anomaly can be identified in devices number II to V, which have a constriction width of $W_{\textrm{QPC}}=100\mbox{-}200\,$nm (Fig.~\ref{Fig2}b-c). The conductance drops below $G_{\textrm{0}}$ but does not reach $0.5G_0$ for wider constrictions like in QPC-I, where $W_{\textrm{QPC}}\approx 250\,$nm (Fig.~\ref{Fig2}a). This behaviour suggests that an interaction between the edge channels is crucial for the appearance of the 0.5 anomaly. The conductance of $e^2/h$ implies the transmission of one channel while the other one is reflected. Preliminary data of the detection of this backscattered state is presented in the supplementary information, Fig.~S3. In that experiment, adjacent voltage probes in a Hall geometry next to a QPC have been used to detect an emerging voltage drop with the QPC entering the 0.5 anomaly regime at $B=0$\,T. Our measurement of $R_\mathrm{xy}$ is consistent with predictions by Landauer-B\"uttiker theory for one reflected helical edge channel.

The conductance in the bulk band gap vanishes for very narrow QPCs as depicted in Fig.~\ref{Fig2}d ($W_{\textrm{QPC}}\approx 25\mbox{-}50\,$nm). In this regime, the transport shows a Coulomb blockade behaviour typical for quantum dots (supplementary information, Fig.~S2). We believe that inter-edge coupling and/or local disorder is responsible for the localization. The suppression of conductance for narrow QPCs sets an experimental upper limit for the wave function width of the edge states. Since we are still able to observe a $G_{\textrm{0}}$ plateau for $W_{\textrm{QPC}}=150\,$nm and no suppression of conductance inside the band gap for $W_{\textrm{QPC}}=100\,$nm, we conclude that the localization of each edge channel has to be smaller than $50\,$nm, in agreement with theory \cite{Papaj2016}. %by \emph{e.g.} Papaj \emph{et al.}   \\
The 0.5 anomaly is observed at large negative gate voltages over a wide voltage range. The gate efficiency in our devices is known from reference Hall bars to be $\Delta n_e / \Delta V \approx 8\mbox{-}10 \times 10^{11}$cm$^{-2}/$V. Therefore, we conclude that the bulk density in the regime of the 0.5 anomaly is strongly $p$-doped ($n_h>1\times 10^{12}$cm$^{-2}$). Bulk transport through the point contact in this regime is suppressed, as will be further discussed below. %(, gate efficency ) gate efficeny estimated from reference Hall bar: $\Delta n_e / \Delta V \approx 8-10 \cdot 10^{11}$cm$^{-2}$).
As shown in Fig.~\ref{Fig1}c, a magnetic field $B \lessapprox 300\,$mT does not influence the 0.5 anomaly. %At higher magnetic fields, Landau level quantization becomes relevant and changes the transport behaviour of the QPC completely.  \\
The QPC conductance of a thinner, but still inverted HgTe quantum well ($d_{\textrm{QW}}\approx 7.0\,$nm$>d_c$) with $W_{\textrm{QPC}}\approx 100\,$nm is shown in Fig.~\ref{Fig2}e. By lowering the gate voltage, first conventional conductance steps are observed. The lowest conductance in this device is around $G_{\textrm{0}}$ indicating the quantum spin Hall regime. We carefully checked that indeed no $0.5$ anomaly is observed in thin quantum wells by studying several QPCs with varying $W_{\textrm{QPC}}$, measured in a large temperature ($25\,$mK up to $10\,$K) and gate voltage range (see Fig.~\ref{Fig2}f).
These findings guide us to the importance of the underlying band structure to identify the mechanism for the $0.5$ anomaly.
%As will be discussed in the next section, the quantum well thickness has impact on the bulk band structure but also, more importantly, drastic consequences for the dispersion and localization of the edge channels.

\section{Band structure calculations}
\begin{figure*}[hbtp]
\includegraphics[width=0.7\textwidth]{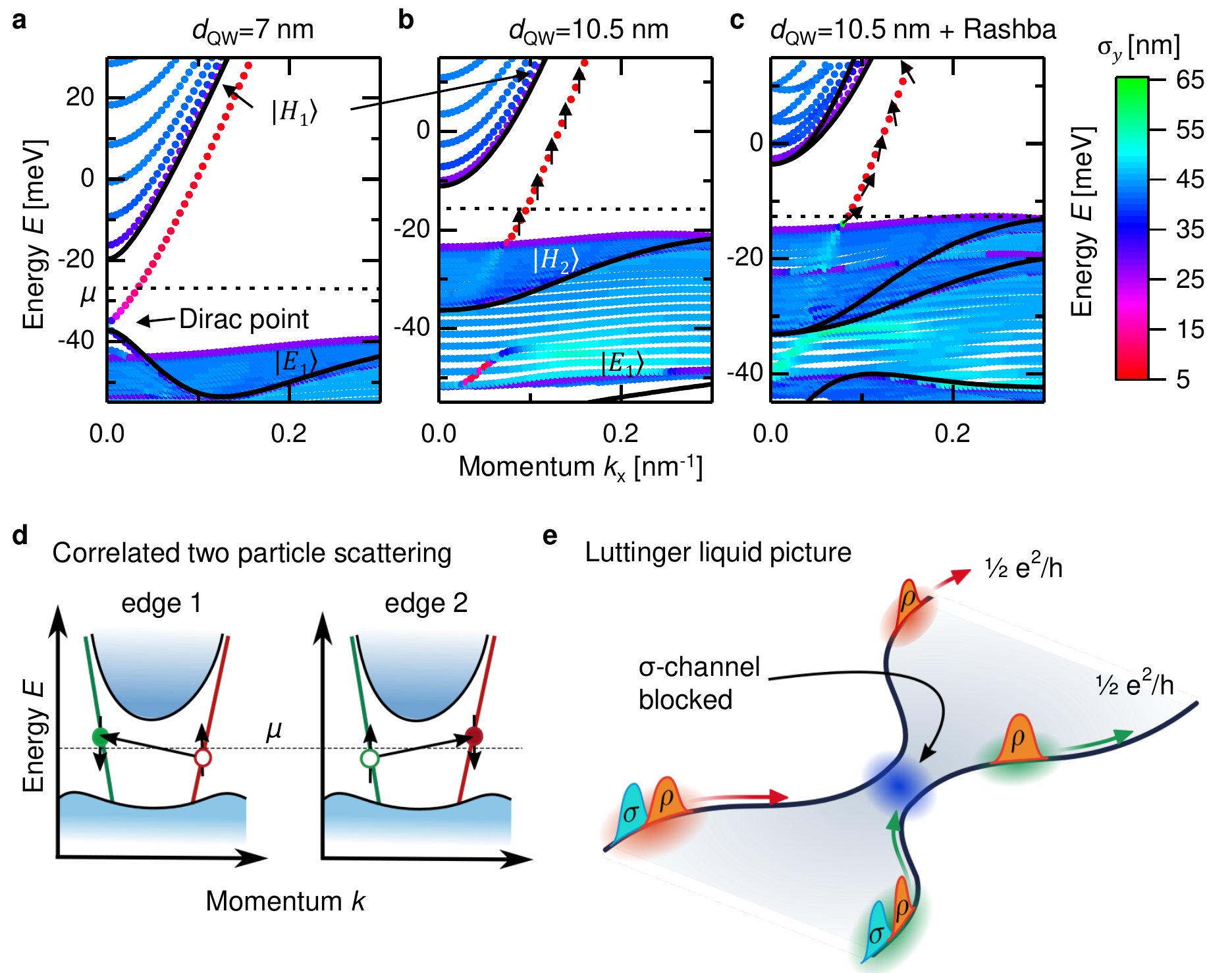}
\caption{\textbf{$k \cdot p$ band structure calculations and illustrations of the scattering process:} \textbf{a-c}, Black lines indicate the bulk band structure. The coloured dots present calculations performed on a finite ribbon with a width of $150\,$nm in the $y$ direction. The colour code shows the wave function standard deviation $\sigma_y$ in the $y$ direction. Small values (red) indicate strongly localized states, large values (blue) bulk like states. Panel \textbf{a} shows the calculated band structure of a $7.0\,$nm quantum well. The Dirac point lies in the bulk band gap. Panel \textbf{b} and \textbf{c} show the band structure for a $10.5\,$nm thick quantum well. Realistic finite electric fields have been applied to the quantum well on the bottom (to simulate iodine doping) and on the top (gate electrode) in panel \textbf{c}, which introduces a Rashba effect as well as a tilting of the spin polarization of the helical edge states away from the normal, as sketched by the arrows. \textbf{d}, Schematic of the correlated scattering process, responsible for the spin gap. \textbf{e}, Illustration of the reduction of conductance in the Tomonaga-Luttinger liquid picture, where $\sigma$ and $\rho$ indicate the bosonic charge and spin fields.}
\label{Fig3}
\end{figure*}

%{\color{cyan} \textit{@Wouter can you take special care for this section?}
%To understand the discrepancy between the two quantum well thicknesses and to get further insights into the transport mechanisms of the QPCs, it is instructive to calculated the band structure of the used quantum wells. This is first done by calculating the infinite 2D structure along the growth axis $z$ of the given HgTe/HgCdTe heterostructure, using the full 8-band Kane $k \cdot p$ model \cite{Novik2005} black lines in Fig.~\ref{Fig3}a)-c). A finite ribbon was simulated in a tight-binding model with hard-wall boundary conditions. This ribbon was chosen to be infinite in $x$-direction and confines the $y$-direction on a scale determined by $W_{\textrm{QPC}}=150\,$nm. The bottom doping and top gating effects of the band structure can be incorporated within a self-consistent approach.}
%The major difference between the two quantum well thicknesses is the band structure.
Using $k \cdot p$ theory based on the eight-band Kane model, we first calculate the bulk band structure of an infinitely wide slab of quantum well material (black curves in Fig.~\ref{Fig3}a-c) \cite{Novik2005}. A more elaborated calculation using a finite width $W_{\textrm{QPC}}=150\,$nm of the system allows us to gain information about the situation inside the QPC constriction (coloured dots in the plots).

The band structure of a quantum well with $d_{\textrm{QW}}= 7\,$nm (Fig.~\ref{Fig3}a) shows the inverted band gap between the $\ket{H_1 \pm}$ and $\ket{E_1 \pm}$ sub-bands as conduction and valence band, respectively. Importantly, the crossing point of the edge channels (Dirac point) lies in the bulk band gap. In contrast, the order of bands in the $10.5\,$nm wide quantum well is rather different (Fig.~\ref{Fig3}b). In this case, the band gap is between the first $\ket{H_1 \pm}$ and second $\ket{H_2 \pm}$ heavy hole sub-band. The $\ket{E_1 \pm}$ sub-band -- still responsible for the band inversion -- lies energetically below the $\ket{H_2 \pm}$ state. Then, the Dirac point is buried deeply in the valence band and the edge states hybridize with the bulk states if they spatially overlap \cite{Skolasinski2017}. However, at the indicated position of the chemical potential in Fig.~\ref{Fig3}c (by the dashed line), the edge states are well localized at the sample edge while the bulk density is already hole dominated. The corresponding edge wave function has a width of approximately $10\,$nm. This value is in qualitative agreement with our observation of unperturbed edge channel transport for QPC widths $W_{\textrm{QPC}}\geq 100\,$nm.

The position of the Dirac point in the valence band and the flat heavy hole bands have several implications for carrier transport. First, lowering the gate voltage in wider quantum wells pushes the chemical potential into the heavy hole $\ket{H_2 \pm}$ bulk sub-bands, where the valence band structure exhibits a camel back-like shape. As a consequence, the Fermi level is pinned at the flat valence band edge. Second, the large Fermi momentum mismatch between valence and conduction band suppresses inter-band transitions and thus also suppresses bulk transport in the $p$-regime. In addition, the separation in momentum space between the edge and bulk states allows their coexistence without hybridization. These arguments explain the range in gate voltage of the quantum spin Hall plateau at $G_{\textrm{0}}$, which is longer than the 'conventional' steps, as well as the suppression of bulk conductance when entering the valence band.
Furthermore, the application of a large negative gate voltage induces a strong Rashba effect. Self-consistent $k\cdot p$ calculations allow us to include the applied electric field and the resulting band structure is shown in Fig.~\ref{Fig3}c \cite{Novik2005}. The dispersion of the bulk bands shows the typical Rashba splitting, while the dispersion of the edge states is not affected. The Rashba coupling does induce an energy dependence of the spin-momentum locking in the edge states as indicated by the tilted  arrows \cite{Schmidt2012, Ortiz2016}.
%but they become generic helical edge channels \emph{i.e.} the spin-momentum locking is now a function of energy as indicated by the tilted arrows. This leads to a breaking of axial spin symmetry for large negative gate voltages \cite{Schmidt2012, Ortiz2016}.
Obviously, a band splitting due to the Rashba coupling alone can not explain a 0.5 anomaly, since it does not break time reversal symmetry \cite{Molenkamp2001}. Hence, we have to take interactions into account.
%Moreover, the valence band edge has almost no dispersion and thus a very low Fermi velocity. In a semiclassical picture an enhancement of $e-e$-interaction is expected, since electrons have more time to interact with each other \cite{Bauer2013}.
\section{Opening of a spin gap due to Coulomb interactions}
In this section, we explain how the emergence of a spin gap generated by correlated two-particle scattering processes can explain the 0.5 anomaly. It is well known that the combination of Rashba spin-orbit coupling and electron-electron interactions at the helical edge can in principle give rise to backscattering (supplementary information, Sec.~III) \cite{Xie2016,Geissler2017}. When both edge channels interact with each other, a variety of two-particle scattering terms are allowed \cite{Teo2009,Sternativo2014,Liu2011}. In general, however, most of these terms are either not relevant in a renormalisation group sense, or do not apply to the constraints set by the band structure in our setup.

As indicated by the $k\cdot p$ calculations, the inverted quantum wells with $d=10.5\,$nm have a Fermi wave vector of $k_F\sim 0.1\, \mathrm{nm}^{-1}$. Backscattering processes, which do not preserve the number of right- and left-moving edge channels, hence, oscillate as a function of space over a scale of $k_F^{-1}$. Since the length of the QPC is of the order of $L\sim 100\,$nm, net effects of these terms should average out.

Following those arguments and assuming (weak) repulsive electron-electron interactions, we show in the supplementary information, Sec.~III, that the most relevant two-particle scattering term can be written has
\begin{align}
\label{Eq:Hs}
H_S=g_s \int_0^L dx [\hat{\chi}^\dagger_{R,+}(x)\hat{\chi}_{L,+}(x)\hat{\chi}^\dagger_{L,-}(x)\hat{\chi}_{R,-}(x)+ \textrm{h.c.}],
\end{align}
where $\hat{\chi}_{\nu,\pm}(x)$ with $\nu\in R, L$ are right- ($R$) and left-moving ($L$) Fermi field operators of upper ($+$) or lower edge ($-$), respectively. Since the spin degree of freedom and the direction of motion are pinned in each helical liquid, we only indicate the direction of motion in Eq.~(\ref{Eq:Hs}) and drop the spin degree of freedom for ease of notation. Evidently, $H_S$ describes a backscattering process between the $(+)$ and $(-)$ edges preserving the number of right- and left movers (see Fig.~\ref{Fig3}d for a schematic).

In our minimal model, introduced in the supplementary information, Sec.~III, Eq.~(\ref{Eq:Hs}) appears due to the combination of Rashba spin-orbit coupling and electron-electron interactions with broken SU(2) symmetry of the spin degree of freedom. The coupling constant $g_s$
\begin{eqnarray}
g_s=\sin^2(\gamma)\frac{g_{2\perp}-g_{4\perp}}{2}
\end{eqnarray}
is found to be directly related to the magnitude of the Rashba coupling strength $\alpha$ via $\gamma=\arctan[\alpha/(\hbar v_F)]$, as well as to the electron-electron interaction processes across the edges parametrized by $g_{2\perp}$ and $g_{4\perp}$. In the presence of strong spin-orbit coupling, SU(2) invariance is broken at the single-particle level. Hence, it makes sense that it remains to be broken in the presence of interactions which implies that $g_{2\perp} \neq g_{4\perp}$.

The Fermi level pinning in the samples with quantum well thickness of $10.5\,$nm, thus, allows the coupling constant $g_s$ to grow, as the electric field and likewise the Rashba coupling is increased. This indicates the importance of the camel back in the bandstructure shown in Fig.~\ref{Fig3} \textbf{c} for the development of a sufficiently large $g_s$.

Using bosonization techniques, we can demonstrate that Eq.~(\ref{Eq:Hs}) acts as a gap to the spin sector\cite{Giamarchi2003}. The effective Hamiltonian reads
\begin{multline}
\label{Eq:Hamiltonian_final}
H_{\mathrm{eff}}=\frac{1}{2\pi}\int_0^L \mathrm{d}x\sum_{\nu=\sigma,\rho}\bigg[\frac{u_{\nu}}{K_{\nu}}\left(\partial_x\phi_{\nu}\right)^2+u_{\nu}K_{\nu}\left(\partial_x\theta_{\nu}\right)^2\bigg]  \\
+\tilde{g}_s\cos(2\sqrt{2}\theta_{\sigma}),
\end{multline}
where $\phi_{\nu}(x),~\theta_{\nu}(x)$ ($\nu\in \rho,\sigma$) describe bosonic fields acting on spin ($\sigma$) and charge sector ($\rho$), $\tilde{g}_s$ is a rescaled version of $g_s$, $u_{\nu}$ represent the normalised velocities and $K_{\nu}$ are the Tomonaga-Luttinger interaction parameters ranging between $0\leq K_{\rho}\leq 1$ and $1\leq K_{\sigma}\leq 1/K_{\rho}$ for a repulsively interacting system. We have dropped the explicit spatial dependence of the bosonic fields for ease of notation. The last term in Eq.~(\ref{Eq:Hamiltonian_final}) -- proportional to $\tilde{g}_s$ -- corresponds to a gap in the spin sector. In the supplementary information, Sec.~III, we explain that (in a mean-field sense) the emergence of the spin gap can be understood as spontaneous time-reversal symmetry breaking.

\section{Experimental consequences of a spin gap}

Usually, spin gaps are not detectable in charge transport experiments of purely one-dimensional systems. However, the strong localization of the single-particle wave functions at the edges of the QPC implies that in the present case the system is by no means a single one-dimensional system, but has to be treated as two spatially separated one-dimensional systems, coupled by Coulomb interactions.

\begin{figure*}[hbtp]
\includegraphics[width=0.7\textwidth]{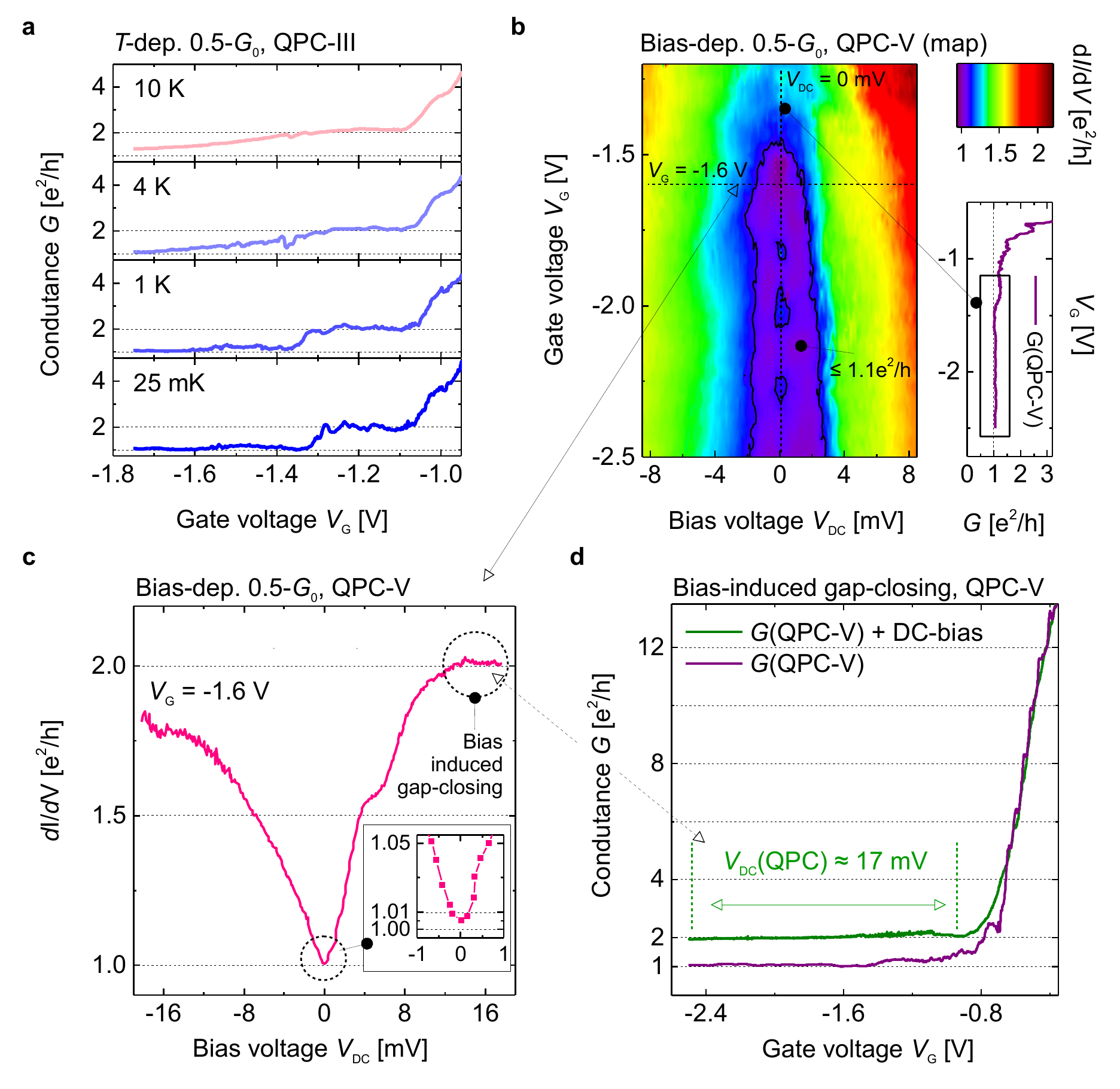}
\caption{\textbf{Temperature and dc bias dependence of the 0.5 anomaly:} \textbf{a}, Temperature dependence of the $2e^2/h$ to $e^2/h$ transition of QPC-III. \textbf{b}, Differential conductance map of QPC-V as a function of bias and gate voltage. The violet area indicates the regime of the 0.5 anomaly. \textbf{c}, Line cut of the bias dependence at $V_{\textrm{G}}=-1.6\,$V. We attribute the asymmetry of the bias dependence to self-gating effects. \textbf{d}, Conductance as a function of gate voltage of QPC-V for zero applied DC bias (violet) and large (green) applied DC bias of $17\,$mV.}
\label{Fig4}
\end{figure*}

Thus, the current operators $j_{\pm}(x)=1/(2\pi)\partial_t (\phi_{\rho}(x)\mp\theta_\sigma(x))$, where the index $\pm$ also relates to different $y$-coordinates, are distinct at the two edges. An electric bias couples to each helical edge state separately. This assumption leads to a reduced
conductance of $G=0.5 G_0$ in the presence of a spin gap (see also Fig.~\ref{Fig3}e). In the absence of the spin gap, we
instead find $G=G_0$ (supplementary information, Sec.~III).

As observable in Fig.~\ref{Fig2}, the fluctuations on top of the 0.5 anomaly plateau are considerably smaller than in the quantum spin Hall regime, where both helical channels are transmitted. In the presence of Eq.~\eqref{Eq:Hs}, our renormalisation group analysis (supplementary information, Sec.~III) indeed predicts a reduced sensitivity to impurity backscattering consistent with this observation. Moreover, we note that the proposed mechanism is not affected by magnetic fields, also consistent with the experiment.

The absence of the 0.5 anomaly in thinner quantum wells can be understood through the lack of Fermi level pinning. In thicker quantum wells (10.5 nm), the application of a strong electric field allows us to generate a sufficiently large Rashba field without substantially affecting the electron density of the edge states. The reason is that the camel back of the valence band has a large density of states at the Fermi energy which gives rise to Fermi level pinning, see the horizontal dashed line in Fig.~\ref{Fig3}c. In contrast, in thinner quantum wells (7 nm), the camel back is far away (in energy) from the Fermi level, see the horizontal dashed line in Fig.~\ref{Fig3}a. Hence, in that case, we are not able to apply strong electric fields without substantially affecting the electron density of the edge states. We argue that the resulting Rashba field, acting on the edge states in the transport regime with conductance $2 e^2/h$, is too small to observe the 0.5 anomaly.

The bias and temperature dependence of the conductance, depicted in Fig.~\ref{Fig4}, helps us to quantify the observed energy scales. As shown in Fig.~\ref{Fig4}a, the 0.5 anomaly is observable up to temperatures of $1.4\,$K. For higher temperatures ($T \geq 4\,$K) the quantization is lost and the conductance increases with increasing temperatures. The range $1\mbox{-}2\,$K as the upper limit to which the quantized plateau is observed sets an energy scale of the spin gap $\Delta E \approx 150\mbox{-}300\,$\textmu eV.
This energy scale is in good agreement with the bias dependence shown in Fig.~\ref{Fig4}b. There, the low ac bias has been superimposed by a dc bias voltage $V_{\textrm{DC}}$. The gate voltage regime in which the $0.5$ anomaly can be observed opens around $V_G=-1.6\,$V. We are able to observe the 0.5 anomaly up to $V_{\textrm{DC}} \approx 200\mbox{-}400\,$\textmu eV (Fig.~\ref{Fig4}c) depending on the gate voltage. A similar estimate can be made for the energy scale set by the length of the QPC $\hbar v_F/L_{\textrm{Gate}}\approx 200\mbox{-}300\,$\textmu eV. The agreement of the magnitudes of all energy and temperature scales is remarkable. We conjecture that they set the typical energy scale required for the development of the 0.5 anomaly. For larger energies, the renormalisation group flow of $g_s$ is stopped too early such that the spin gap can not develop.

Increasing the applied bias voltage further, the conductance increases beyond the 0.5 anomaly and a second step like plateau is visible around $\approx 0.8 G_0$ (Fig.~\ref{Fig4}c). We conjecture that this feature is related to the 0.7 anomaly commonly observed in conventional QPCs. The emergence of this conventional 0.7-like signature is in qualitative agreement with the explanation given in Refs.~\cite{Bauer2013, Lunde2009, sloggett2008} for GaAs based structures. In these articles, electron-electron interactions at the bottom of the last sub-band suppress the conductance below $G_0$. In our case, the 0.7 feature occurs where the applied bias becomes large enough to touch the bottom of the interaction induced gap. Depending on the device, we are also sometimes able to identify a 0.7 feature as a function of gate voltage (see Fig.~\ref{Fig2}b). Increasing the bias even further closes the interaction induced gap and the conduction saturates at $G_{\textrm{0}}$, \emph{i.e.} two unperturbed edge channels are now perfectly transmitted through the QPC over a large range of gate voltage (see Figs.~\ref{Fig4}c and d).

Several other mechanisms  might explain the $0.5$ anomaly in QPCs or nanowires. These mechanisms include helical edge reconstruction \cite{Wang2017}, the formation of a Wigner crystal \cite{Matveev2004}, or hyperfine interactions \cite{Hsu2018}. However, given the importance of the camel back in the valence band for our observation of the 0.5 anomaly, we believe that the mechanism presented here is the most plausible one.  At the same time, we note (and discuss this more extensively in the supplementary information) that one can imagine another relevant mechanism, in particular, the helical edge reconstruction proposed in Ref.~\cite{Wang2017}, that shares many common ingredients to our mechanism -- like strong spin-orbit coupling, electron-electron interactions, and confinement. Hence, it is likely that the two mechanisms are related to each other (from a more fundamental point of view). Importantly, the explanation of the $0.5$ anomaly relies in any case on the spontaneous breaking of time-reversal symmetry by interactions.

\section{Summary \& outlook}
To conclude, we have presented the realization and operation of a QPC in a two-dimensional topological insulator. The conductance as a function of applied gate voltage saturates on a robust and reproducible $0.5G_0$ plateau. Investigations of this 0.5 anomaly for various QPC channel widths, combined with the fact that the 0.5 feature is linked to a certain quantum well thickness, gives a hint to the importance of the underlying band structure. Especially, the difference between a Dirac point in the band gap and one buried in the valence band guides us to a scattering term, which implies the opening of a spin gap. The 0.5 anomaly yields an effectively spin-polarized current, which may find applications in spintronics. Furthermore, the results could be important for the detection of Majorana bound states since the identified mechanism might be related to the observation of the $4\pi$-periodic Josephson current in our HgTe Josephson junctions in the absence of an explicit time reversal symmetry breaking mechanism \cite{Bocquillon2016}. Combining a topological QPC with superconductors is envisaged to enable the creation and manipulation of Majorana bound states and parafermions \cite{Li2016,Fleckenstein2018}.
% Lastly, it its worth noting that the realization of a QSHI QPC is a first experimental step into a completely new field -- helical electron quantum optics \cite{Ferraro2014c}.
% The description within the Luttinger picture and the presence of $e-e$-correlations allows experimental studies in the field of correlated helical 1D systems and fractional topological excitations. Lastly, an effective low energy model similar to the famous Bernevig-Hughes-Zhang model which describes the dispersion of the edge channels in the buried case (effectively all HgTe quantum wells with a thickness $d_{\textrm{QW}}>8\,$nm is lacking \cite{Bernevig2006}.

\subsection*{Acknowledgements}
We like to thank E. Bocquillon, T. Borzenko, Y. Gefen, C. Gould, V. Hock, P. Leubner, and Y. Meir for fruitful discussions.
We acknowledge financial support by the DFG (SPP1666 and SFB1170 ”ToCoTronics”), the ENB Graduate school on ”Topological Insulators”, the EU ERC-AG Program (project 4-TOPS), and the W\"urzburg-Dresden Cluster of Excellence on Complexity and Topology in Quantum Matter (EXC 2147,
project-id 39085490). C.F. acknowledges support from the Studienstiftung des Deutschen Volkes.
\subsection*{Author Contributions}
J.S. prepared the samples and performed the experiments. V.M. and P.S. contributed to implement the fabrication process, S.S. and J.W. helped to do the measurements. J.K. supervised the sample fabrication. J.W. guided the experiments. L.L. grew the material. W.B. provided the code for the band structure calculations. C.F., N.T.Z. and B.T. developed the theoretic model. H.B. and L.W.M. planned the project and design of the experiment. All authors participated in the analysis of the data, led by J.S. and J.W. All authors jointly wrote the manuscript, led by J.W.
\subsection*{Competing interests}
The authors declare that they have no competing financial interests.
\subsection*{Data availability}
The data that support the plots within this paper and other findings of this study are available from the corresponding author upon reasonable request.

\clearpage
\newpage
\widetext

\newcommand{\beginsupplement}{
        \setcounter{table}{0}
        \renewcommand{\thetable}{S\arabic{table}}
        \setcounter{figure}{0}
        \renewcommand{\thefigure}{S\arabic{figure}}
				\setcounter{equation}{0}
				\renewcommand{\theequation}{S\arabic{equation}}
				\setcounter{page}{1}
				\renewcommand{\thepage}{S-\arabic{page}}
				\setcounter{section}{0}
				\renewcommand{\thesection}{\arabic{section}}
     }
		
\beginsupplement

\begin{center}
\textbf{\large Supplementary Information for}\\
\vspace{0.5cm}
\textit{\Large \textbf{Interacting topological edge channels}}
\end{center}

\section{Methods}
The quantum point contacts (QPCs) were fabricated from HgTe quantum wells, epitaxially grown on CdZnTe substrates. An iodine doping layer $70\,$nm below the quantum well is used to increase the electron density and mobility of the equilibrium reservoirs. This layer decreases the parasitic lead resistance. Reference Hall-bars are used to determine the density $n_e$ and mobility $\mu$ of the wafer material, which are presented in Tab.~\ref{tab1}. %The transport through the QPC is ballistic since $l_{\textrm{mfp}}\approx 4 \mu\textrm{m}  \ll L_{\textrm{QPC}}$.

The first step of the lithographic fabrication of the QPCs is the definition of the equilibrium reservoirs without the QPC constriction using electron beam lithography ($2.5\,$kV acceleration voltage) and wet etching with an aqueous solution of $\textrm{KI}:\textrm{I}_2:\textrm{HBr}$ \cite{Bendias2018}. In order to have precise control over the width $W_{\textrm{QPC}}$, the constrictions are etched in a second step using the same etchant. In a third step all outer HgTe areas, which would else shunt the bond pads, are etched. This three step mesa process allows the reproducible fabrication of QPCs with dimensions of $W_{\textrm{QPC}}$ as small as $25\,$nm and a length $L$ of roughly $500\,$nm, well below $l_{\textrm{mfp}}$. The advantage of wet etching in contrast to dry etching techniques is that the edges are not affected by local doping, which reduces the mobility especially in small structures drastically. Narrow top-gate electrodes are realized using electron beam lithography with an acceleration voltage of $6.5\,$kV. The length of the gate is approximately  $L_{\textrm{Gate}} \approx 200\mbox{-}300\,$nm. A low temperature atomic layer deposition process ($T<40^\circ$C) is used to deposit $15\,$nm of HfO$_2$ dielectric and the electrodes are metallized with Ti/Au. AuGe/Au ohmic contacts using optical lithography methods are structured far away from the QPC constriction to assure full energy relaxation in the HgTe reservoirs ($> 10 \times l_{\textrm{mfp}}$). Special care is taken that during all process steps the temperature never exceeds $80^\circ$C.
The number of modes in the QPC is given by $N=\sqrt{2n_e 4 W_{\textrm{QPC}}/\pi}$ and the density $n_e$ in the constriction is controlled by a voltage applied on the narrow gate electrode.

Most of the transport measurements were performed in Helium-4-cryostats at $1.4$\,K using standard four point low frequency low bias lock-in techniques. Complementary measurements were conducted in a dilution refrigerator with a base temperature of $25\,$mK.

\begin{table}[htbp]
\begin{center}
\begin{tabular}{|c|c|c|c|c|}
\hline QPC Nr. & $W_{\textrm{QPC}}/$nm & $d_{\textrm{QW}}\,/$nm & $n_e(0\,\textrm{V})/(10^{11}\,\textrm{cm}^{-2})$ & $\mu(0\,\textrm{V})/(10^{5}\,\textrm{cm}^2 \textrm{V}^{-1}\textrm{s}^{-1})$ \\
\hline I & 250 & 10.5 & 5.9 & 3.2 \\
\hline  II & 200 & 10.5 & 5.9 & 3.2  \\
\hline  III & 150 & 10.5 & 5.9 & 3.2 \\
\hline  IV &  100 & 10.5 & 5.9 & 3.2 \\
\hline  V &  100 & 10.5 & 5.9 & 3.2 \\
\hline  VI & 25 & 10.5 & 5.9 & 3.2 \\
\hline  VII & 100 & 7.0 & 5.2 & 2.7 \\
\hline  QPC-bar & 200  & 10.5 & 5.8  & 2.0 \\
\hline
\end{tabular}
\label{tab1}
\end{center}

\caption{Summary of parameters of samples mentioned in the main text. Density $n_e$ and electron mobility $\mu$ were obtained from reference Hall-bar measurements.}
\end{table}

\newpage
\section{Complementary experimental data}
\begin{figure}[hbtp]
\includegraphics[width=\linewidth]{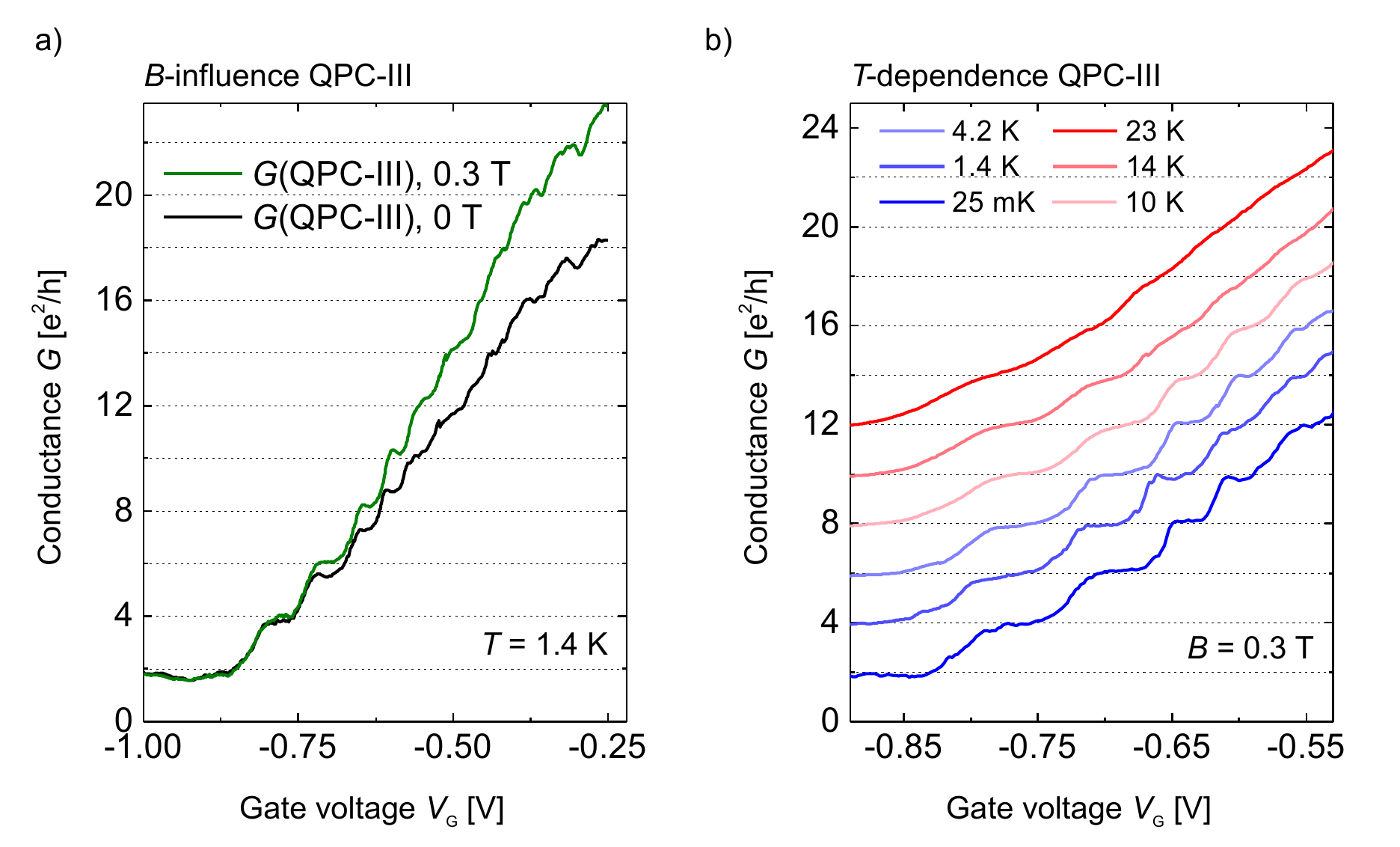}
\caption{a) A small magnetic field $B$ reduces the parasitic serial resistance and leads to a sequence of well quantized plateaus. b) Temperature dependence of the conventional plateaus measured at $0.3\,$T. From the temperature dependence of the plateaus we are able to estimate an energy splitting of the sub-bands to be around $\Delta E\ \lesssim 4 k_B T\approx 4.8\,$meV. The curves are offset vertically for clarity. %$2e^2/h$ to $e^2/h$ transition of QPC-III for QPC-II. b) Differential conductance map of QPC-V as a function of bias and gate voltage. The violet area indicates the regime of the 0.5 anomaly. c) Line cut of the bias dependence at $V_{\textrm{G}}=-1.6\,$V. The asymmetry of the bias dependence is attributed to self-gating effects. d) Conductance as a function gate voltage of QPC-IV for zero applied DC bias (violet) and large ($17\,$meV) applied DC bias.
}
\label{SupFig1}
\end{figure}
\newpage

\begin{figure}[hbtp]
\includegraphics[width=\linewidth]{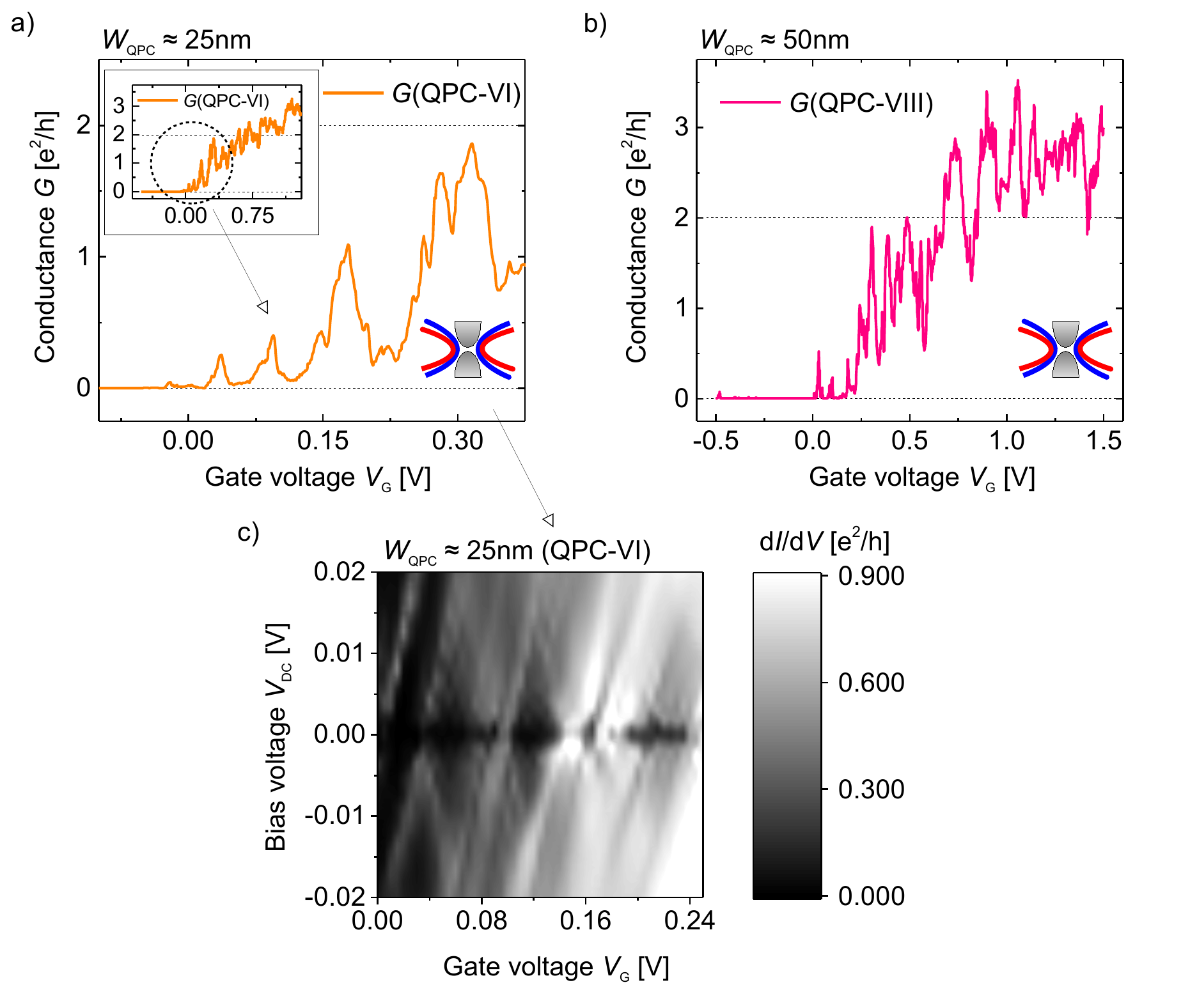}
\caption{Complementary data for narrow QPCs. a) and b) show the conductance as a function of gate voltage for a QPC with a width $W_{\textrm{QPC}}$ of $25\,$nm and $50\,$nm. c) Bias voltage dependence of the conductance. Coulomb blockade behaviour like in a quantum dot is observed indicating the single electron tunnel limit regime of the QPC.   %$2e^2/h$ to $e^2/h$ transition of QPC-III for QPC-II. b) Differential conductance map of QPC-V as a function of bias and gate voltage. The violet area indicates the regime of the 0.5 anomaly. c) Line cut of the bias dependence at $V_{\textrm{G}}=-1.6\,$V. The asymmetry of the bias dependence is attributed to self-gating effects. d) Conductance as a function gate voltage of QPC-IV for zero applied DC bias (violet) and large ($17\,$meV) applied DC bias.
}
\label{SupFig2}
\end{figure}

\newpage

\begin{figure}[hbtp]
\includegraphics[width=\linewidth]{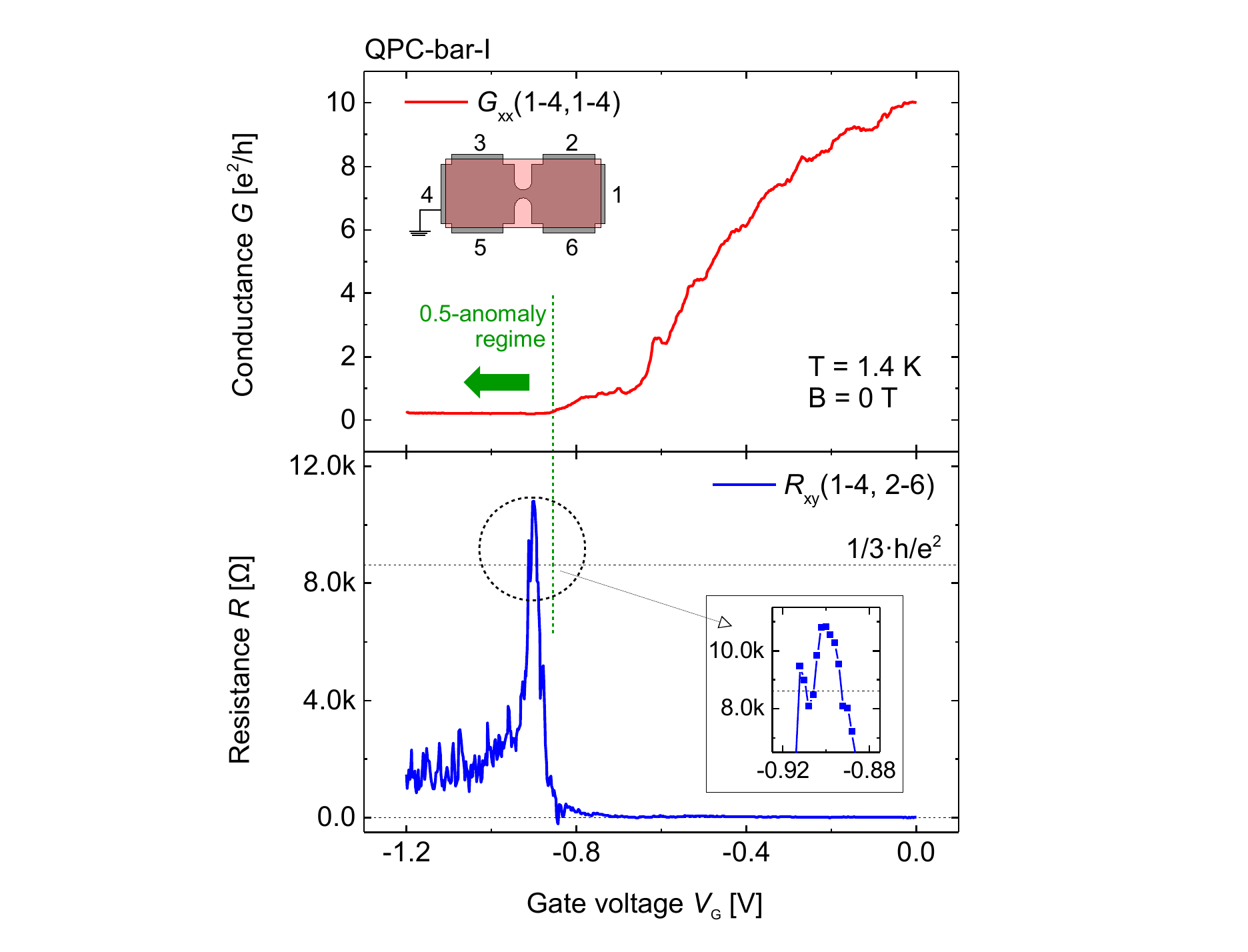}
\caption{Preliminary data of a Hall bar with a QPC in the centre as depicted in the inset. This device allows the detection of the backscattered edge channel by measuring the longitudinal and transversal resistance in parallel. Clear step like structures, resembling the behaviour of the QPC presented in the main text, can be identified. The conductance values do not match the expected steps due to a gate voltage dependent non-linear background resistance. Nevertheless, we are able to identify the regime of the 0.5 anomaly. The transversal resistance (lower panel) is zero up to the transition of to the 0.5 anomaly. It then enhances suddenly and the measured value is close to the value expected from Landau-B\"{u}ttiker calculations for one scattered edge channel (dashed horizontal line). A finite Hall voltage at zero applied magnetic field is only possible if edge channels are responsible for the phenomena.     %$2e^2/h$ to $e^2/h$ transition of QPC-III for QPC-II. b) Differential conductance map of QPC-V as a function of bias and gate voltage. The violet area indicates the regime of the 0.5 anomaly. c) Line cut of the bias dependence at $V_{\textrm{G}}=-1.6\,$V. The asymmetry of the bias dependence is attributed to self-gating effects. d) Conductance as a function gate voltage of QPC-IV for zero applied DC bias (violet) and large ($17\,$meV) applied DC bias.
}
\label{SupFig3}
\end{figure}
The multi-terminal device is patterned with an integral top gate. This choice causes the effective gate action within the lead areas closest to the QPC to be more sensitive as compared to the QPC area itself. The reason is that the thicknesses of the deposited HfO$_2$ differ for both regions due of lithographic processing (lead areas: 45 cycles HfO$_2$; QPC area: 90 cycles HfO$_2$). This difference reduces the stability and thus the measurability of reflected helical edge channels for increasing negative values of $V_\mathrm{G}$. We believe that the leads enter the p-regime earlier (referring to increasing negative values of $V_\mathrm{G}$) than the QPC. Our future lithographic development will therefore focus on the realization and implementation of three separated top gates. This design should enable us to independently manipulate and tune the different sample areas.

\newpage

%\begin{figure}[hbtp]
%\includegraphics[width=\linewidth]{Figs/supplement_04_biasmap_21_transition}
%\caption{Additional bias voltage dependent data for QPC-II.
%}
%\label{SupFig4}
%\end{figure}

\newpage

\section{Theoretical derivation of the scattering mechanism}
\label{sec:theo}

In this section, we provide a detailed derivation of the mechanism we employ to interpret the $0.5$ anomaly. In Sec. \ref{sec:1}, we discuss the emergence of gap-opening terms when Rashba spin-orbit coupling (SOC) is present in the QPC and derive the effective Hamiltonian. Subsequently, in Sec. \ref{sec2}, we compute the conductance induced from the effective Hamiltonian in a linear response picture. Lastly, in Sec. \ref{sec3}, we discuss the influence of impurity induced backscattering in the presence of the spin-gap phase.
\subsection{Interaction induced backscattering with Rashba spin-orbit coupling}
\label{sec:1}
At the helical edge of two-dimensional topological insulators, time reversal invariant Rashba SOC appears as a consequence of externally applied electric fields or electric fields due to geometric boundaries. Both cases are relevant for QPCs formed in quantum spin Hall systems. In the absence of electron-electron interactions, Rashba SOC alone cannot generate backscattering within a single helical edge state. However, in the presence of electron-electron interactions, the interplay between them and Rashba SOC can result in backscattering under certain conditions \cite{Geissler2017}.

We now explain the emergence of a particular type of correlated two-particle scattering if both forward scattering (due to electron-electron interactions) and Rashba SOC are present. In the QPC under consideration, Rashba SOC can be formulated by three contributions
\begin{eqnarray}
H_{\mathrm{R}}=H_{1\mathrm{R}}+H_{2\mathrm{R}}+H_{12\mathrm{R}},
\end{eqnarray}
describing the Rashba SOC within each edge ($H_{1\mathrm{R}}$ and $H_{2\mathrm{R}}$) and across the edges ($H_{12\mathrm{R}}$). Each term  connects a spin-flip mechanism with the momentum operator
\begin{eqnarray}
\label{Eq:Rashba1}
H_{1\mathrm{R}}\!\!&=&\!\!\int\!\!\mathrm{d}x\alpha\!\left[\hat{\psi}_{L1}^{\dagger}(x)i\hat{p}_x\hat{\psi}_{R1}(x)\!-\!\hat{\psi}_{R1}^{\dagger}(x)i\hat{p}_x\hat{\psi}_{L1}(x)\!\right]\!\!,\\
\label{Eq:Rashba2}
H_{2\mathrm{R}}\!\!&=&\!\!\int\!\!\mathrm{d}x\alpha\!\left[\hat{\psi}_{L2}^{\dagger}(x)i\hat{p}_x\hat{\psi}_{R2}(x)\!-\!\hat{\psi}_{R2}^{\dagger}(x)i\hat{p}_x\hat{\psi}_{L2}(x)\!\right]\!\!,\\
\label{Eq:Rashba12}
H_{12\mathrm{R}}\!\!&=&\!\!\int\!\!\mathrm{d}x\tilde{\alpha}\!\left[\!\hat{\psi}_{R2}^{\dagger}(x)i\hat{p}_x\hat{\psi}_{R1}(x)\!-\!\hat{\psi}_{R1}^{\dagger}(x)i\hat{p}_x\hat{\psi}_{R2}(x)\right]\nonumber\\
&+&\!\!\int\!\!\mathrm{d}x\tilde{\alpha}\!\left[\!\hat{\psi}_{L1}^{\dagger}(x)i\hat{p}_x\hat{\psi}_{L2}(x)\!-\!\hat{\psi}_{L2}^{\dagger}(x)i\hat{p}_x\hat{\psi}_{L1}(x)\right]\!\!,~~~~~~
\end{eqnarray}
where the prefactors $\alpha$ and $\tilde{\alpha}$ characterize the coupling strength, which might be different within each edge as compared to across the edges.
All three terms are time-reversal invariant. Together with the kinetic part of the Hamiltonian, we can reorganize Eqs. (\ref{Eq:Rashba1}-\ref{Eq:Rashba12}) into
\begin{equation}
\label{Eq:Hkin+HR}
H_{\mathrm{kin}}\!+\!H_{\mathrm{R}}\!=\!\int \mathrm{d}x\hat{\Psi}^{\dagger}(x)\!\left[\hat{p}_x(\alpha\tau_z\sigma_y\!\!+\!\!\tilde{\alpha}\tau_y\sigma_z)\!+\!v_F\hat{p}_x\tau_0\sigma_z\right]\!\hat{\Psi}(x)
\end{equation}
with the basis $\hat{\Psi}(x)=(\hat{\psi}_{R1}(x),\hat{\psi}_{L1}(x),\hat{\psi}_{R2}(x),\hat{\psi}_{L2}(x))^T$ and $\tau_i$, $\sigma_i$ being Pauli-matrices acting on edge-, spin-space respectively. For ease of notation, we have dropped the spin index in this description. Due to helicity, spin and direction of motion are strongly coupled to each other and opposite for the two edges. A valid choice for the spin degree of freedom could, for instance, be that the $R$-movers of edge 1 and the $L$-movers of edge 2 have spin $\uparrow$, whereas the $L$-movers of edge 1 and the $R$-movers of edge 2 have spin $\downarrow$.

In the next step, we diagonalize the Hamiltonian in Eq. (\ref{Eq:Hkin+HR}). This can be done by the following unitary transformation $\hat{\Psi}(x)=U\hat{\chi}(x)$ with $U=e^{(i\pi/4)\tau_x\sigma_0}e^{(i/2)\gamma \tau_z\sigma_x}$, where $\gamma=\arctan(\alpha/v_F)$. Note that $U$ acts non-trivially on edge- and spin-space. The transformed Hamiltonian $\tilde{H}_{\mathrm{kin}}+\tilde{H}_{\mathrm{R}}=U(H_{\mathrm{kin}}+H_{\mathrm{R}})U^\dagger$ becomes
\begin{equation}
\label{Eq:Hkin+HR1}
\tilde{H}_{\mathrm{kin}}\!+\!\tilde{H}_{\mathrm{R}}\!=\!=\!\int\!\!\!\mathrm{d}x\hat{\chi}^{\dagger}(x)\!\!\left[\frac{v_+}{2}(\tau_z\!+\!\tau_0)\sigma_z\!+\!\frac{v_-}{2}(\tau_0\!-\!\tau_z)\sigma_z\right]\!\hat{p}_x\!\hat{\chi}(x)
\end{equation}
with the new basis $\hat{\chi}(x)=(\hat{\chi}_{R+}(x),\hat{\chi}_{L+}(x),\hat{\chi}_{R-}(x),\hat{\chi}_{L-}(x))^T$, where the fields $\hat{\chi}_{R/L,\pm}(x)$ correspond to annihilation operators of a $R/L$-moving fermion in edge $\pm$. After the transformation, the $R/L$-moving excitations do not carry the same spin degrees of freedom anymore as the $R/L$-movers before the transformation. Instead, they carry linear combinations thereof. Likewise, the $\pm$ edge degree of freedom corresponds to a linear combination of the original $1/2$ edge degree of freedom. The two edge sectors ($\pm$) have a different Fermi velocity
\begin{equation}
v_{\pm}\!=\!v_F\!\!\left[\frac{1}{\cos(\gamma)}\pm \tan(\tilde{\gamma})\right].
\end{equation}
For simplicity, we neglect the inter-edge Rashba SOC from now on by choosing $\tilde{\alpha}=0$ (i.e. $v_-=v_+$). Then, the transformation matrix $U$ takes the compact form
\begin{eqnarray}
\label{Eq:trafo}
U=e^{(i/2)\gamma \tau_z\sigma_x}.
\end{eqnarray}
Note that this transformation is diagonal in edge space (because the Pauli matrix $\tau_z$ is diagonal). Hence, under the assumption $\tilde{\alpha}=0$, the $\pm$ index for the edges are identical to the original $1/2$ index.

The above analysis implies that, including Rashba SOC, no backscattering appears for the case of free (non-interacting) Dirac fermions, as the resulting theory can again be mapped on a theory of free Dirac fermions. However, when electron-electron interactions are present, we have to include forward scattering (density-density) interaction terms
\begin{eqnarray}
\label{Eq:g2}
H_{g_2}&=&\int \mathrm{d}x g_2\left[\hat{n}_{R1}(x)\hat{n}_{L1}(x)\!+\!\hat{n}_{R2}(x)\hat{n}_{L2}(x)\right],\\
\label{Eq:g4}
H_{g_{4\perp}}&=&\int\mathrm{d}x g_{4\perp}\left[\hat{n}_{R1}(x)\hat{n}_{R2}(x)+\hat{n}_{L1}(x)\hat{n}_{L2}(x)\right],\\
\label{Eq:g2perp}
H_{g2\perp}&=& \int\mathrm{d}x g_{2\perp} \left[\hat{n}_{R1}(x)\hat{n}_{L2}(x)+\hat{n}_{R2}(x)\hat{n}_{L1}(x)\right]
\end{eqnarray}
with the density operators $\hat{n}_{\nu,1/2}(x)=\hat{\psi}_{\nu 1/2}^{\dagger}(x)\hat{\psi}_{\nu 1/2}(x)$ of edge $1$, $2$, respectively. Note that these terms are still written in the original basis $\hat{\Psi}(x)$. When transformed into the new fermions $\hat{\chi}_{\nu}(x)$, the density operators $\hat{n}_{\nu,1/2}(x)$ imply backscattering terms $\sim \hat{\chi}_{R+}^{\dagger}(x)\hat{\chi}_{L+}(x)+\mathrm{h.c.}$. Thus, the product of density operators in Eqs.~(\ref{Eq:g2}-\ref{Eq:g2perp}) necessarily leads to correlated two-particle backscattering terms in the new basis. Since we put $\tilde{\alpha}=0$ for simplicity, correlated scattering between the edges is only generated by Eqs.~(\ref{Eq:g4}) and (\ref{Eq:g2perp}).

For the regime of large chemical potential in both edges, i.e. $k_{F+},k_{F-}\gg L^{-1}$ (with $L$ being the length of the QPC and $k_{F\pm}$ the Fermi wave-vector in edge $\pm$) and (weak) repulsive interactions, the only term that is relevant in renormalization group (RG) sense and preserves the number of right- and left-movers is given by \cite{Gritsev2005, Cheng2011, Sedlmayr2013, Fleckenstein2018}
\begin{equation}
\label{Eq:HsSuppMat}
\mathcal{H}_{\mathrm{s}}(x)=g_s\big[\hat{\chi}_{R+}^{\dagger}(x)\hat{\chi}_{L+}(x)\hat{\chi}_{L-}^{\dagger}(x)\hat{\chi}_{R-}(x)+\mathrm{h.c.}\big].
\end{equation}
Since this Hamiltonian contains field operators of both kind of edge states ($\pm$), it only oscillates according to the chemical potential imbalance between the edges $\delta k_F=k_{F+}-k_{F-}$. For Eq. (\ref{Eq:HsSuppMat}) to be significant, we thus require $\delta k_F\ll L^{-1}$. This is a reasonable assumption because $k_{F+} \approx k_{F-}$.

In our effective theory, the coupling constant $g_s$ is directly related to the strength of electron-electron interactions as well as the rotation angle of Rashba SOC $\gamma$. From an expansion of Eq. (\ref{Eq:g4},\ref{Eq:g2perp}) using Eq. (\ref{Eq:trafo}) we obtain
\begin{equation}
g_s=\sin^2(\gamma)\frac{g_{2\perp}-g_{4\perp}}{2}.
\end{equation}
The coupling constant, thus, vanishes in the case of SU(2)-symmetric interactions with $g_{2\perp}=g_{4\perp}$. In the QPC formed in the quantum spin Hall insulator, SU(2) invariance is broken at the single particle level (due to strong spin-momentum locking at the two separate edges). Hence, it makes sense that SU(2) invariance is also broken in the presence of interactions. In other words, the combination of Rashba SOC (in the bulk, within each edge, and between the edges) and Coulomb interactions implies that $g_{2\perp} \neq g_{4\perp}$.

If axial spin symmetry is broken, again by some kind of SOC, then the term written in Eq.~(\ref{Eq:HsSuppMat}) is in principle allowed (better to say: not forbidden) by symmetry arguments. By that reasoning, we could have postulated it (without the careful derivation presented above) from the start \cite{Schmidt2012,Ortiz2016}. Another way to argue for a finite $g_s$ is related to the edge reconstruction mechanism proposed by Wang, Meir, and Gefen \cite{Wang2017}. These authors have developed a model that predicts a spatial separation of the right- and the left-movers at a single edge of the quantum spin Hall system (due to a smooth confinement potential in combination with Coulomb interactions). This edge reconstruction implies spontaneous symmetry breaking of time-reversal symmetry and naturally leads to a finite $g_s$ because the spatial distance between the involved densities in the $g_{2\perp}$-term and the $g_{4\perp}$-term would be different.

Let us think about a possible mean-field treatment of the Hamiltonian (\ref{Eq:HsSuppMat}). We could postulate (by hand) a plausible choice of the mean-field potential such as $M_{RL,\pm}(x)=\langle \hat{\chi}_{R\pm}^{\dagger}(x)\hat{\chi}_{L\pm}(x) \rangle$ in which the expectation value is taken with respect to some symmetry broken ground state. In fact, this choice is formally equivalent to a ferromagnetic ordering in $x$-direction within the channel $\pm$. Evidently, this mean-field potential implies spontaneous breaking of time-reversal symmetry. A finite value of $M_{RL,\pm}(x)$ corresponds to a magnetic gap in the channel $\pm$. How does the corresponding mean-field Hamiltonian look like? Defining $\tilde{M}_{RL,\pm}(x)=M_{RL,\pm}(x) g_s$, we can write the two possible choices of the mean-field Hamiltonians in the following form
\begin{eqnarray}
\mathcal{H}^{\mathrm{MF}+}_{\mathrm{s}}(x)&=&\tilde{M}_{RL,+}(x) \hat{\chi}_{L-}^{\dagger}(x)\hat{\chi}_{R-}(x)+\mathrm{h.c.}\, , \label{hmf1} \\
\mathcal{H}^{\mathrm{MF}-}_{\mathrm{s}}(x)&=&\tilde{M}_{RL,-}(x) \hat{\chi}_{R+}^{\dagger}(x)\hat{\chi}_{L+}(x)+\mathrm{h.c.} \, . \label{hmf2}
\end{eqnarray}
Let us concentrate on the former one, Eq.~(\ref{hmf1}), for concreteness: A finite value of $\tilde{M}_{RL,+}(x)$ implies a magnetic gap in the $+$ channel. Moreover, $\mathcal{H}^{\mathrm{MF}+}_{\mathrm{s}}(x)$ describes a backscattering in the $-$ channel, due to the appearance of the operator product $\hat{\chi}_{L-}^{\dagger}(x)\hat{\chi}_{R-}(x)$. Hence, such a mean-field treatment can be used to describe backscattering across the QPC but it does it in the two channels ($\pm$) simultaneously. The reason is the correlated scattering in the two channels described by the original Hamiltonian (\ref{Eq:HsSuppMat}). Therefore, we argue that the mean-field treatment (outlined above) cannot describe the 0.5 anomaly properly because it is unable to predict a stable region of conductance $e^2/h$. Below we show that a proper linear response theory is able to make this prediction considering the correlated pair scattering (\ref{Eq:HsSuppMat}) beyond a mean-field description.

In the next step, we bosonize the theory in terms of the new fermions $\hat{\chi}_{\eta}(x)$ using the standard bosonization identity
\begin{eqnarray}
\hat{\chi}_{r\nu}(x)=\frac{\mathcal{F}_{r,\nu}}{\sqrt{2\pi\alpha}}e^{-\frac{i}{\sqrt{2}}\big[r\phi_{\rho}(x)-\theta_{\rho}(x)+\nu(r\phi_{\sigma}(x)-\theta_{\sigma}(x))\big]},
\end{eqnarray}
where $r=R,L=+,-$ and $\nu=+,-$. $\mathcal{F}_{r,\nu}$ are Klein factors lowering the number of fermions by one.
The conjugate bosonic fields $\phi_{\rho/\sigma}(x),~\theta_{\rho/\sigma}(x)$ are linear combinations of the bosonic fields of the edges $+$ and $-$: $\phi_{\rho}=1/\sqrt{2}(\phi_+(x)+\phi_-(x)),~\phi_{\sigma}=1/\sqrt{2}(\theta_-(x)-\theta_+(x)),~\theta_{\rho}=1/\sqrt{2}(\theta_+(x)+\theta_-(x)),~\theta_{\sigma}=1/\sqrt{2}(\phi_-(x)-\phi_+(x))$. Note that the fields with index $\rho$ can be viewed as charge excitations and the fields with index $\sigma$ as spin excitations. The resulting Hamiltonian, composed of the kinetic term, the Rashba SOC, and the interaction terms, then reads
\begin{equation}
\label{Eq:Heffective}
H = \frac{1}{2\pi}\int \mathrm{d}x\sum_{\nu=\sigma,\rho}\bigg[\frac{u_{\nu}}{K_{\nu}}\left(\partial_x\phi_{\nu}(x)\right)^2+u_{\nu}K_{\nu}\left(\partial_x\theta_{\nu}(x)\right)^2\bigg] + \tilde{g}_s\cos(2\sqrt{2}\theta_{\sigma}(x)),
\end{equation}
where $\tilde{g}_s=g_s/(2\pi^2\alpha^2)$, $u_{\nu}$ are renormalized velocities and $K_{\nu}$ are the Luttinger liquid interaction parameters. The latter remain form invariant to the case of vanishing Rashba SOC but with renormalized coupling constants
\begin{eqnarray}
K_{\rho}&=&\frac{\sqrt{1-\bar{g}_2+\bar{g}_{4}-\bar{g}_{2\perp}+\bar{g}_{4\perp}}}{\sqrt{1+\bar{g}_2+\bar{g}_4+\bar{g}_{2\perp}+\bar{g}_{4\perp}}},\\
K_{\sigma}&=&\frac{\sqrt{1+\bar{g}_2+\bar{g}_4-\bar{g}_{2\perp}-\bar{g}_{4\perp}}}{\sqrt{1-\tilde{g}_2+\bar{g}_4+\tilde{g}_{2\perp}-\tilde{g}_{4\perp}}},
\end{eqnarray}
where $\bar{g}_{\nu}=g_{\nu}/(2\pi v)$ and
\begin{eqnarray}
\bar{g}_2&=&g_2\eta+4 g_4 \xi,~~~~~~~\bar{g}_4=g_4 \eta+ g_2 \xi,\\
\bar{g}_{2\perp}&=&g_{2\perp} \eta +2g_{4\perp} \xi,~~ \bar{g}_{4\perp}=g_{4{\perp}} \eta + 2g_{2\perp} \xi~
\end{eqnarray}
with $\eta=\cos(\gamma/2)^4+\sin(\gamma/2)^4$ and $\xi=\cos(\gamma/2)^2\sin(\gamma/2)^2$.

The term proportional to $\tilde{g}_s$ in Eq.~(\ref{Eq:Heffective}) is called a mass term in field theory. It gives rise to a gap in the spin sector $\sigma$ whereas the charge sector $\rho$ is still described by a free boson.
\
\subsection{Conductance}
\label{sec2}
We now calculate the impact of the previously derived mass term on the conductance. In particular, the system we investigate is formulated by two one-dimensional systems (helical Luttinger liquids) that are independent in the range $x\leq 0$ and $x\geq L$, while they are coupled to each other in the region of the QPC ($0\leq x \leq L$). The assumption of a step-like variation of the coupling is valid under the following conditions: The Fermi wave length $\lambda_F$ should be much smaller than the smoothing length $L_s$ that describes the build-up of the mass term as the QPC is formed which again should be much smaller than the length $L$ of the QPC, i.e. we need the following hierarchy of length scales $\lambda_F \ll L_s \ll L$ \cite{Dolcini2005}.

As the QPC is directly contacted in the experimental setup, the applied electrostatic potential difference produces an electric field which is mainly concentrated in the constricted region of the QPC. In a linear response ansatz, the current is thus given by
\begin{eqnarray}
\label{Eq:Current}
I(x,t) = \sum_{j=1,2} I_j(x,t) = \sum_{j=1,2} \int_0^L \mathrm{d}x' \int \frac{\mathrm{d}\omega}{2\pi} e^{-i\omega t} \sigma_{j}(x,x',\omega) E_j(x',\omega)
\end{eqnarray}
with Fourier transformed electric fields $E_{1/2}(x',\omega)$, conductances $\sigma_{1/2}(x,x',\omega)$ and currents $I_{1/2}$ in the one-dimensional system $1$, $2$, respectively. Note that we use the index $1,2$ again (not $+,-$ anymore) for the two helical Luttinger liquids. In the absence of inter-edge Rashba SOC, however, the two choices of indexes are identical.

\begin{figure}
\centering
\includegraphics[scale=0.5]{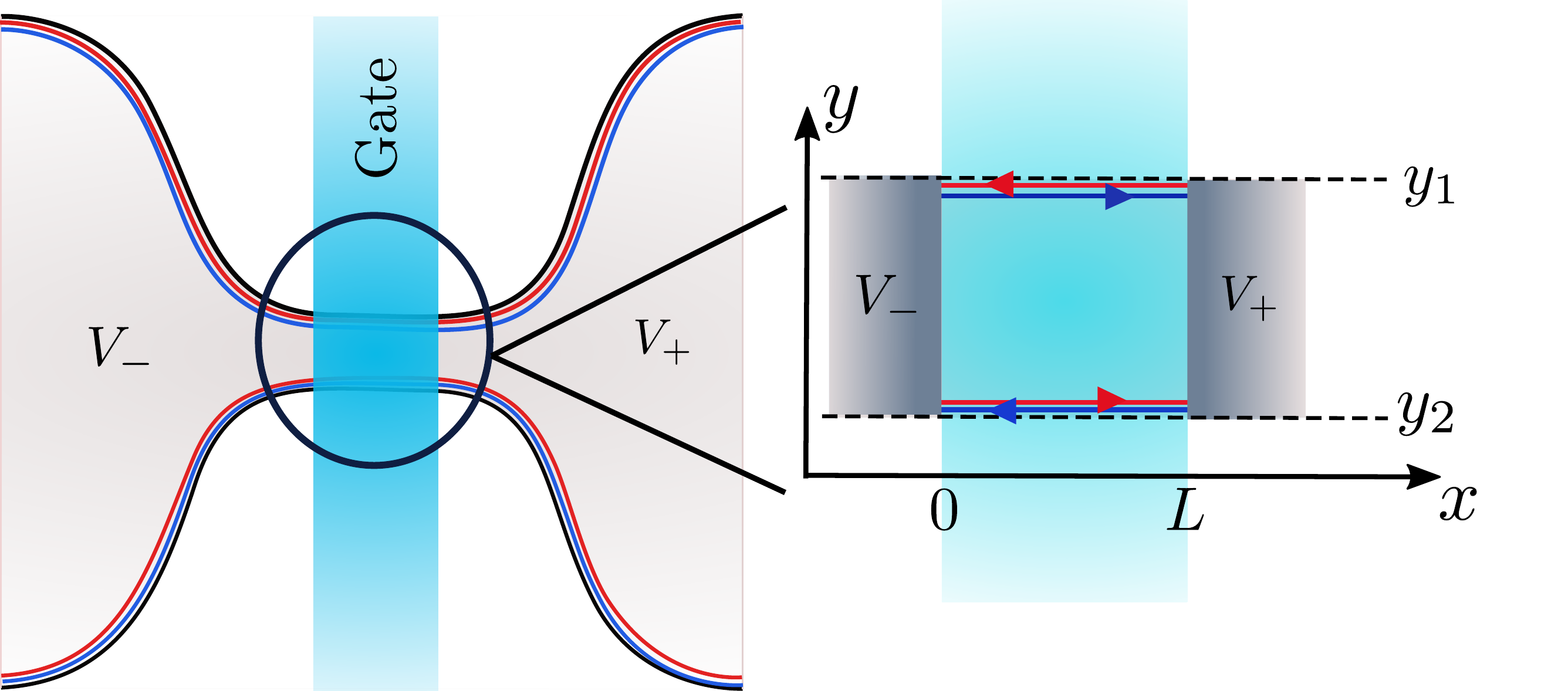}
\caption{Schematic of the setup: The width of the QPC is much larger than the decay length of the single-particle wave functions.}
\label{Fig:QPC1}
\end{figure}

As the electric field couples to the current, the conductance is related to the current-current correlation function in linear response. For the result of the calculation, the geometrical setup is of significant importance. As demonstrated by $k\cdot p$ calculations, the edge states decay on length scales $\xi_l\ll d=y_1-y_2$ (Fig. \ref{Fig:QPC1}), where $d$ is the width of the QPC. Thus, the physical system acts as two (coupled) one-dimensional systems and the index $1,2$ of the currents can also be related to the $y$-coordinate.
%
% In a true quantum wire, the electron denstiy and the current are determiend by the charge sector ($\rho$) of the Luttinger liquid $1$ and $2$ (as everything is a truely 1D model)
%\begin{equation}
%2\pi\hat{n}_w(x)= \hat{n}_1(x)+\hat{n}_2(x),~~\hat{j}_w(x)\!=\partial_t/(2\pi)(\hat{n}_1(x)+\hat{n}_2(x)).
%\end{equation}
The separation of the edge states implies that the electric field at $y=y_1$ (see Fig.~\ref{Fig:QPC1}), coined $E_1(x,\omega)$ in Eq.~(\ref{Eq:Current}), can only couple to the current operator of helical liquid $1$ (the upper one), while at $y=y_2$ it correspondingly couples to the current operator of helical liquid $2$ (the lower one).

With this assumption, it follows that the conductances $\sigma_{1}(x,x',\omega)$ and $\sigma_{2}(x,x',\omega)$ are given by
\begin{eqnarray}
\label{Eq:Propagator1}
\sigma_{i}(x,x',\omega)=-e^2 \frac{\omega}{\pi} G_{ii,\omega}(x,x')
\end{eqnarray}
with
\begin{eqnarray}
G_{ii,\omega}(x,x')=\int_0^{\beta}\frac{d\tau}{(2\pi)^2}\langle \phi_i(x,\tau),\phi_i(x',0) \rangle_0 e^{-i\omega \tau},
\end{eqnarray}
where the expectation value is taken with respect to the unperturbed system \cite{Maslov1995}. The propagator $G_{\omega}(x,x')$ is determined by the action, which we derive from the effective Hamiltonian (\ref{Eq:Heffective}) (under the assumption that $\tilde{g}_s(x)$ is only finite for $0 \leq x \leq L$)
\begin{equation}
S\!=-\!\int\!\!\mathrm{d}t\!\!\int\!\!\mathrm{d}x\!\frac{1}{2}\Phi^T(x,t)\big[L_0(x)\partial_t^2\!-\!\partial_x L_1(x) \partial_x\!-\!M(x)\big] \Phi(x,t),
\end{equation}
where $\Phi(x,t)=(\phi_1(x,t),\phi_2(x,t))^T$. $L_0(x)$ and $L_1(x)$ are $2\times 2$ matrices containing all information about the couplings between sector $1$ and $2$ by the interactions. In general, they can be parameterized as
\begin{equation}
L_0(x)=\frac{1}{a'b'\!-\!c'(x)^2}\!\begin{pmatrix}
b'\!& \!c'(x)\\
c'(x)\! & \!a' \\
\end{pmatrix}~~,~L_1(x)=\begin{pmatrix}
a \!& \!c(x)\! \\
c(x)\! & \!b\! \\
\end{pmatrix}.
\end{equation}
Notice that in the case of vanishing Rashba SOC between the edges we have $a=b$, $a'=b'$ and furthermore $a=u_0/K_0$, $a'=u_0 K_0$ with the Luttinger parameters $K_0$ and $u_0$ of isolated interacting helical edges. %Moreover, it should be mentioned that Rashba SOC does not direenter $L_0(x)$ and $L_1(x)$, which is reasonable since it should not change the equation of motion in the case of vanishing interactions.
The parameters $c(x)$ and $ c'(x)$ describe the interactions across the two helical edges. Their spatial dependence is modelled by $c(x)=c z(x)$, $c'(x)=c' z(x)$ with
\begin{eqnarray}
z(x)=\bigg\lbrace \begin{matrix}
1&,&~~ 0\leq x \leq L,\\
0&,&~~\mathrm{else}
\end{matrix}.
\end{eqnarray}
The mass $M(x)\equiv M z(x)$ is the large $\tilde{g}_s$ approximation of Eq. (\ref{Eq:Heffective}). Under the assumption that the corresponding field only takes small deviations around the value minimizing the cosine potential $\theta_{\sigma}(x)\rightarrow \theta_{\sigma,0}+\theta_{\sigma}(x)$, this yields
\begin{eqnarray}
\tilde{g}_s(x) \cos[2\phi_2(x,t)-2\phi_1(x,t)]\! \simeq \! -\!\tilde{g}_s(x)\!+\!\frac{\tilde{g}_s(x)}{2}[2\phi_2(x,t)\!-\! 2\phi_1(x,t)]^2\nonumber
=-\tilde{g}_s(x)\!+\!\Phi(x,t)^TM(x)\Phi(x,t)
\end{eqnarray}
with
\begin{eqnarray}
M(x)=2 \tilde{g}_s(x)\begin{pmatrix}
~1 & -1\\
-1 & ~1\\
\end{pmatrix}.
\end{eqnarray}
Applying standard field theoretical methods, we obtain the propagator satisfying the equation
\begin{equation}
\label{Eq:Porpagator2}
\big[\! L_0(x)\omega^2\!-\!\partial_x L_1(x)\partial_x\!-\! M(x)\big]G_{\omega}(x,x')=-\delta(x-x')\mathbb{1}_{2\times 2},
\end{equation}
where $G_{\omega}(x,x')$ contains all possible correlations
\begin{eqnarray}
G_{\omega}(x,x')=\begin{pmatrix}
G_{11,\omega}(x,x') & G_{12.\omega}(x,x')\\
G_{21,\omega}(x,x') & G_{22,\omega}(x,x')\\
\end{pmatrix}.
\end{eqnarray}
%(The minus on the right hand side of Eq. (\ref{Eq:Porpagator2}) arises from the derivation of the propagator using source terms in the action.)
To solve Eq. (\ref{Eq:Porpagator2}), we first search for the eigenfunctions of the homogeneous problem
\begin{eqnarray}
\label{Eq:scattering}
\big[\! L_0(x)\omega^2\!-\!\partial_x L_1(x)\partial_x\!-\! M(x)\big]\Psi_i(x,\omega)=0.
\end{eqnarray}
Since we have a $2\times 2$ second order differential equation, we expect to find four distinct eigenfunctions. The problem is similar to a scattering problem, with a constant scattering potential in region $0\leq x \leq L$. Any eigenfunction, as well as its derivative, must be continuous at each interface. Valid eigenfunctions for $x\leq 0$ are given by
\begin{eqnarray}
\Psi_1(x,\omega)&=& \chi_1(x,\omega)+g_{12} \chi_2(x,\omega)+ g_{14} \chi_4(x,\omega),\\
\Psi_2(x,\omega)&=& \chi_3(x,\omega)+g_{22} \chi_2(x,\omega)+ g_{24} \chi_4(x,\omega),\\
\Psi_3(x,\omega)&=& g_{32} \chi_2(x,\omega)+ g_{34} \chi_4(x,\omega),\\
\Psi_4(x,\omega)&=& g_{42} \chi_2(x,\omega)+ g_{44} \chi_4(x,\omega)
\end{eqnarray}
with
\begin{eqnarray}
\chi_1(x,\omega)&=& (1,0)^T \exp[-\omega \gamma x],\\
\chi_2(x,\omega)&=& (1,0)^T \exp[\omega \gamma x],\\
\chi_3(x,\omega)&=& (0,1)^T \exp[-\omega \gamma x],\\
\chi_4(x,\omega)&=& (0,1)^T \exp[\omega \gamma x],
\end{eqnarray}
where $\gamma=1/\sqrt{a a'}$. The coefficients $g_{ij}$ are fixed by continuity conditions at the interfaces. The Green function of the second order differential equation (\ref{Eq:Porpagator2}) has to be a continuous function as $x\rightarrow x'$, but undergoes a jump in the derivative
\begin{eqnarray}
\label{Eq:propagator3a}
\lim_{\epsilon\rightarrow 0} G_{\omega}(x,x')\big\vert_{x=x'-\epsilon}^{x=x'+\epsilon}&=&0,\\
\label{Eq:propagator3b}
\lim_{\epsilon\rightarrow 0} L_1(x)\partial_x G_{\omega}(x,x')\big\vert_{x=x'-\epsilon}^{x=x'+\epsilon}&=&\mathbb{1}_{2\times2}.
\end{eqnarray}
Furthermore, the asymptotic behaviour of the Green function has to obey
\begin{equation}
\label{Eq:propagator4}
\lim_{x\rightarrow -\infty}G_{\omega}(x,x')=0,
\end{equation}
which corresponds to outgoing wave boundary conditions in real time formulation. This suggests the ansatz
\begin{eqnarray}
G_{\omega}(x,x')=[\Psi_{3}(x,\omega)A_3^T+\Psi_4(x,\omega)A_4^T]\theta(x'-x)+ [\Psi_{1}(x,\omega)A_1^T+\Psi_2(x,\omega)A_2^T]\theta(x-x').
\end{eqnarray}
Eqs. (\ref{Eq:propagator3a}-\ref{Eq:propagator3b}) determine the vectors $A_i$ that depend on $x'$ and $\omega$. Eventually, we obtain
\begin{eqnarray}
G_{11,\omega}(x,x')&=&\!-\frac{K_0}{2\omega}\bigg[ e^{(x'\!-\! x)\gamma
\omega}\left(1\!+\! g_{12}e^{2x\omega \gamma}\right)\theta(x\!-\! x')
+e^{(x\!-\! x')\omega\gamma}(1\!+\! g_{12}e^{2x'\omega\gamma})\theta(x'\!-\! x)\bigg],\\
G_{22,\omega}(x,x')&=&\!-\frac{K_0}{2\omega}\bigg[ e^{(x'\!-\! x)\gamma
\omega}\left(1\!+\! g_{24}e^{2x\omega \gamma}\right)\theta(x\!-\! x')
+e^{(x\!-\! x')\omega\gamma}(1\!+\! g_{24}e^{2x'\omega\gamma})\theta(x'\!-\! x)\bigg],\\
G_{12,\omega}(x,x')&=&\! -\frac{K_0 g_{22}}{2\omega}e^{(x+x')\omega\gamma},\\
G_{21,\omega}(x,x')&=&\! -\frac{K_0 g_{14}}{2\omega}e^{(x+x')\omega\gamma}.
\end{eqnarray}
We are interested in the dc limit $\omega\rightarrow 0$ of the conductance, given in Eq. (\ref{Eq:Propagator1}). With $h=2\pi$ ($\hbar=1$), this yields
\begin{eqnarray}
\label{Eq:propagator5}
\lim_{\omega\rightarrow 0}\sigma(x,x',\omega)=\lim_{\omega\rightarrow 0}
\begin{pmatrix}
\sigma_{11}(x,x',\omega) & \sigma_{12}(x,x',\omega) \\
\sigma_{21}(x,x',\omega) & \sigma_{22}(x,x',\omega) \\
\end{pmatrix}=
\frac{e^2K_0}{h}\begin{pmatrix}
1+g_{12} & g_{22}\\
g_{14} & 1+g_{24}\\
\end{pmatrix}.
\end{eqnarray}
Eq. (\ref{Eq:propagator5}) determines the propagator for $x\leq 0$ to be a constant function of $x$ and $x'$ in the $\omega\rightarrow 0$ limit. This is a rather general property, since the $x$ and $x'$ dependence is merged in exponentials $\sim\exp[ax\omega+bx'\omega]$ with $\{a,b\}\in\mathbb{C}$. Thus, we can conclude that the propagator is also a constant function of $x$ and $x'$ for $0\leq x,x'\leq L$. With the continuity of the propagator at each interface, it follows that the propagator takes the same value everywhere.

The constants $g_{12}$, $g_{22}$, $g_{14}$ and $g_{24}$ are derived as the solution of the scattering problem, set by Eq. (\ref{Eq:scattering}). We want to distinguish two cases: (i) the massive case and (ii) the mass-less case. For both cases, we take the limit $\omega\rightarrow 0$ of the obtained solution. This yields the results presented in Tab. \ref{Tab:1}.
With these results, we can now derive the conductance of the system in the two cases. From Eq. (\ref{Eq:Current}), we obtain
\begin{eqnarray}
\mathrm{(i)}~~~~I_1+I_2&=&\frac{K_0 e^2}{h}\bigg[\frac{1}{2}+\frac{1}{2}\bigg]\int_0^Ldx' E(x')\nonumber\\
&=& \frac{K_0 e^2}{h}(V_--V_+),\\
\mathrm{(ii)}~~~~I_1+I_2&=&\frac{K_0 e^2}{h}\bigg[1+1\bigg]\int_0^Ldx' E(x')\nonumber\\
&=& \frac{2K_0 e^2}{h}(V_--V_+).
\end{eqnarray}
The electrostatic potentials $V_\pm$ are schematically shown in Fig.~\ref{Fig:QPC1}. When we assume the QPC to be contacted by (weakly interacting) Fermi-liquid leads, which is a reasonable assumption, we can put $K_0\sim 1$ \cite{Maslov1995, Safi1995, Ponomarenko1995}. Thus,  we find a conductance of $G=e^2/h$ for case (i) and $G=2e^2/h$ for case (ii).
\begin{table}[h]
\centering
\begin{tabular}{|c||c|c|}
\hline
 & $g_s=0$ & $g_s$ large\\
\hline\hline
$g_{12}$ & 0 & $-1/2$ \\
$g_{24}$ & 0 & $-1/2$ \\
$g_{22}$ & 0 & $1/2$ \\
$g_{14}$ & 0 & $1/2$ \\
\hline
\end{tabular}
\caption{Values of the scattering amplitudes in the $\omega\rightarrow 0$ limit for two cases.}
\label{Tab:1}
\end{table}
\subsection{Suppressed backscattering in the massive case}
\label{sec3}
Scattering off disorder in the QPC, this can be modelled as
\begin{eqnarray}
\hat{H}_{dis}=\int\mathrm{d}x  V(x)\bigg[\hat{\psi}_{R,\uparrow}^{\dagger}(x)\hat{\psi}_{L,\uparrow}(x)+\hat{\psi}_{R,\downarrow}^{\dagger}(x)\hat{\psi}_{L,\downarrow}(x)\bigg]+\mathrm{h.c.},
\end{eqnarray}
where $V(x)\sim \delta(x)$ is used for a single impurity. In the bosonized form, we obtain
\begin{eqnarray}
\label{Eq:Hdis}
\hat{H}_{dis}=\int\mathrm{d}x V(x)\cos[\sqrt{2\pi K_{\rho}}\phi_{\rho}(x)-2k_Fx] \cos[\sqrt{2\pi K_{\sigma}}\phi_{\sigma}(x)].
\end{eqnarray}
Note that if Eq.~(\ref{Eq:Hdis}) only applies to a single point-like scatterer, the oscillations proportional to $2k_Fx$ do not matter. Then, the impact of Eq.~(\ref{Eq:Hdis}) is not suppressed by large chemical potentials. In presence of impurities, however, the conductance is expected to deviate from the quantized value.

\begin{figure}
\includegraphics[scale=0.5]{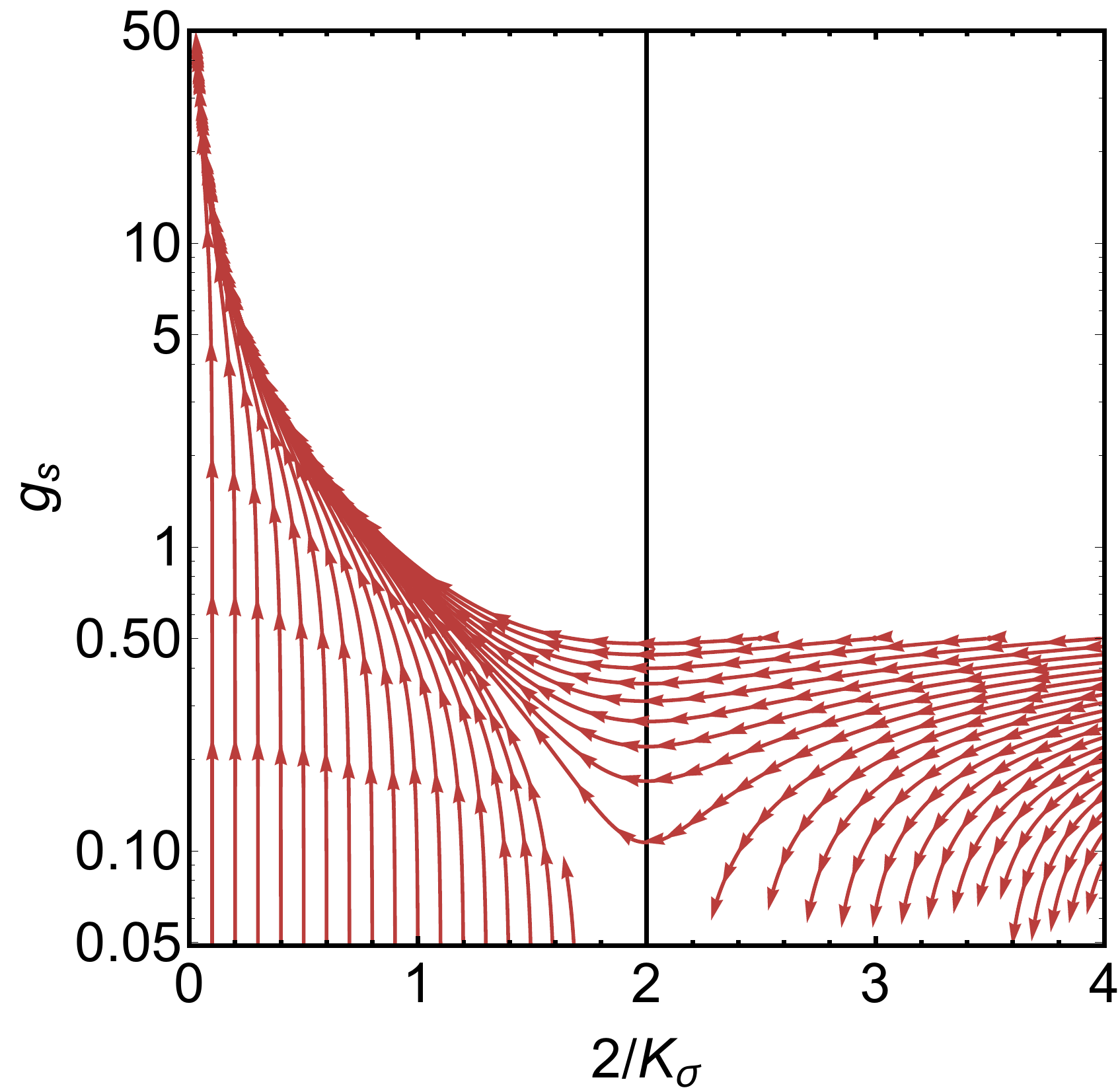}
\caption{RG flow of Eq. (\ref{Eq:Heffective}) up to second order.}
\label{Fig.2}
\end{figure}

To understand the suppression of impurity scattering when the system is gapped by Eq. (\ref{Eq:Heffective}), we need a refined RG analysis. In first order RG, the coupling constant (in our case $\tilde{g}_s$) gets renormalized. Hence, we can enter a regime, where the specific term is relevant and flows to strong coupling. Second order RG also renormalizes the scaling dimension (in our case this is related to $K_{\sigma}$). Applying RG up to second order for Eq. (\ref{Eq:Heffective}), this yields the flow depicted in Fig. \ref{Fig.2}.
When $\tilde{g}_s$ is initially small, then Eq. (\ref{Eq:Heffective}) flows to strong coupling for $K_{\sigma}>1$. In this case, we simultaneously also obtain $K_{\sigma}\rightarrow \infty$.
\newline
\newline
On the other hand,  in first order RG, Eq. (\ref{Eq:Hdis}) is relevant and flows to strong coupling provided the following relation is satisfied
\begin{equation}
\label{Eq:rel1}
K_{\sigma}+K_{\rho}<2.
\end{equation}
Assuming that Eq. (\ref{Eq:Hdis}) does not significantly influence the flow of $K_{\sigma}$ (produced by Eq. (\ref{Eq:Heffective})), Eq. (\ref{Eq:rel1}) can not be satisfied since $K_{\sigma}\rightarrow \infty$. Hence, $\hat{H}_{dis}$ represents a RG irrelevant perturbation.
In conclusion, we expect to find a better quantization of the remaining $1e^2/h$ conductance plateau when the system is gapped by Eq. (\ref{Eq:Heffective}).

\newpage
\section{Estimate of $K_L$}
$K_{\rho}$ can be estimated considering long range Coulomb interactions \cite{Teo2009}
\begin{align}
K_{\rho}=\left[1+\frac{2 e^2}{\pi \epsilon_0 \epsilon_\mathrm{r} \hbar v_\mathrm{F}} \ln \left(\frac{7.1 d}{\xi+0.8 w}\right)\right]^{-1/2}
\end{align}
with $\epsilon_\mathrm{r}\approx 10-20$, $v_\mathrm{F} \approx 1 \cdot 10^5\,$m/s, $d\approx 20\,$ the distance from the edge channel to the gate electrode, $\xi \approx 10-40\,$nm the evanescent decay length of the edge channels and $w\approx 10.5\,$nm the quantum well thickness. Given the uncertainties for the material parameters values between $K_L \approx 0.4-0.8$ can be obtained and thus \emph{posteriori} justifying the approach using the Luttinger liquid formalism with weak repulsive interaction.

\section{Discussion of other potential $0.5G_0$ mechanisms}

Wang, Meir, and Gefen \cite{Wang2017} have recently proposed an interesting mechanism considering a more realistic edge potential due to smooth confinement. This mechanism promotes edge reconstruction, and as a consequence the spatial separation of helical edge states. If those states are brought together in a QPC, the wave functions of the 'inner' pair of edge channels start to overlap earlier than the outer pair, leading to a selective backscattering and a conductance of $0.5G_0$ in the QPC. The theory is based on effective models, which do not properly take into account the dispersion of the edge channels in the case of a buried Dirac cone. We have carefully checked using band structure calculations that the position of the Dirac point for the $10.5\,$nm quantum well can not be moved into the gap by either electrostatic gating (right panel of Fig.\,3 in the main text) or by narrowing the ribbon geometry, which mimics the overlap of edge channels due to spatial separation. Nevertheless, a refined version of this theory might be able to explain the $0.5$-anomaly. In fact, the basic ingredients of the theory by Wang, Meir, and Gefen are: (i) helical edge states (due to strong spin-orbit coupling), (ii) Coulomb interactions, and (iii) smooth confinement. Hence, these basic ingredients are similar to the ones of our theoretical model described in Sec.~\ref{sec:theo}.

Matveev \cite{Matveev2004} discussed the breakdown of spin-charge separation in a one-dimensional quantum wire for the Wigner-crystal regime. The mechanism considers a nanowire where the two sectors of the Hamiltonian -- spin and charge -- are characterized by different energy scales. It would thus lead to a finite temperature window for the 0.5 anomaly. In fact, at low temperatures, the Wigner crystal is a perfect conductor, with conductance $G_0$.

Another proposal considers the hyper-fine interaction of electrons with the nuclear spins which can spontaneously break TR symmetry and lead to an $k_F$ independent partial gap \cite{Hsu2018, Aseev2017}. Because of the low non-zero nuclear spin in HgTe, we expect the gap to be one order of magnitude smaller than in systems based on GaAs. Importantly, this effect should be independent of the QW thickness, which is inconsistent with our data.

\bibliographystyle{apsrev4-1}
\bibliography{QPC_library_arxiv}

\end{document}